\documentclass[reprint,floatfix, aps,prc, twocolumn,notitlepage,
nofootinbib,nobibnotes,
amsmath,amssymb,preprintnumbers]{revtex4-1}
\usepackage{adjustbox}
\usepackage{graphicx,xcolor}
\usepackage{physics}
\usepackage{enumitem}   
\usepackage{dcolumn}
\usepackage{hyperref}

\usepackage[normalem]{ulem}
\usepackage{mathtools}
\usepackage{bm}
\usepackage{subfigure} 
\usepackage{xcolor}
\usepackage{float}
\usepackage{placeins}
\usepackage{tabularx}
\usepackage{url}
\usepackage{latexsym}
\usepackage{epsf}
\usepackage{etoolbox}
\usepackage{cleveref}

\makeatletter
\appto\abstract{%
  \let\latexlist\list
  \def\list{\edef\keeprightskip{\the\rightskip}\latexlist}%
  \patchcmd\latexlist{\ignorespaces}{\rightskip\keeprightskip\ignorespaces}{}{}%
}
\makeatother

\newcommand{\bs}{\boldsymbol}

\begin{document}
\title{Bayesian analysis of a (3+1)D hybrid approach with initial conditions from hadronic transport}
\author{Niklas G\"otz$^{1,2}$, Iurii Karpenko$^{3}$  and Hannah Elfner$^{4,1,2,5}$}
\affiliation{$^1$Goethe University Frankfurt, Department of Physics, Institute for Theoretical Physics, 60438 Frankfurt, Germany}
\affiliation{$^2$Frankfurt Institute for Advanced Studies,  60438
Frankfurt am Main, Germany}
\affiliation{$^3$Faculty of Nuclear Sciences and Physical Engineering, Czech Technical University in Prague,
 11519 Prague 1, Czech Republic}
\affiliation{$^4$GSI Helmholtzzentrum f\"ur Schwerionenforschung,  64291
Darmstadt, Germany}
\affiliation{$^5$Helmholtz Research Academy Hesse for FAIR (HFHF), GSI Helmholtz Center,
Campus Frankfurt,  60438 Frankfurt am Main, Germany}

\date{\today}
\begin{abstract}
\begin{description}
\item[Background] The SMASH-vHLLE-hybrid model integrates the SMASH transport code for the hadronic phase and the vHLLE hydrodynamic model for the QGP phase in heavy-ion collisions.  Bayesian analysis offers a robust statistical method to constrain model parameters of a hybrid approach and evaluate uncertainties by comparing predictions with experimental data. Traditional modelling often relies on parametric assumptions for initial conditions, which may limit the ability to constrain the parameter space due to overfitting.
\item[Purpose] This study aims to apply statistical learning, specifically Bayesian inference, to the (3+1)D SMASH-vHLLE-hybrid model using initial conditions generated by the SMASH transport code itself, with the objective of constraining model parameters and gaining deeper insight on the temperature and baryochemical potential dependence of both the shear and the bulk viscosity.
\item[Method] This study is performed in the hybrid approach SMASH-vHLLE, composed of the hadronic transport approach SMASH and the (3+1)D viscous hydrodynamic code vHLLE. A Bayesian framework is employed, utilizing Markov Chain Monte Carlo (MCMC) sampling to explore the parameter space. The analysis compares model predictions against experimental observables in the range of $\sqrt{s_{NN}}= 7.7 - 200 GeV$, including particle yields, momentum and flow coefficients both at midrapidity as well as in forward and backward direction. 
 \item[Results] We find that the SMASH-vHLLE-hybrid framework, using hadronic initial conditions for Au+Au collisions at different beam energies, can reproduce a variety of experimental observables at midrapidity and forward/backward rapidities. Notably, the preferred posterior distribution suggests a near-vanishing specific shear viscosity in the high-temperature QGP phase, combined with moderate-to-large bulk viscosity around the phase transition region, although the constraints on baryochemical potential dependence are weak.
\item[Conclusions] Our findings reveal that a hadronic initial condition constrains the evolution more strictly at intermediate energies, making parameters such as the hydrodynamic onset time highly sensitive. Intriguingly, the extracted shear viscosity differs substantially from previous Bayesian analyses, motivating further systematic studies with higher-statistics data sets and refined modeling assumptions.
\end{description}
 
\end{abstract}

\maketitle
\section{Introduction}
The study of Quark-Gluon Plasma (QGP) characterization has remained a central focus in high-energy nuclear physics, particularly within the experimental frameworks provided by facilities such as the Relativistic Heavy Ion Collider (RHIC) at Brookhaven National Laboratory and the Large Hadron Collider (LHC) at CERN~\cite{Achenbach:2023pba, Arslandok:2023utm}. These experiments enable a detailed exploration of the QGP, a state of matter where quarks and gluons, the fundamental constituents of protons and neutrons, are deconfined due to extreme temperatures and densities. A crucial component of this investigation is the RHIC Beam Energy Scan (BES) program, which systematically varies the center-of-mass energy of colliding ions to probe the QCD phase diagram over a wide range of temperature and baryon chemical potential~\cite{Caines:2009yu, Mohanty:2011nm, Mitchell:2012mx, Odyniec:2015iaa}. The BES program is particularly significant as it seeks to identify the transition between hadronic matter and the QGP, and to locate a potential critical point and first-order phase boundaries within the QCD phase diagram, thereby offering insights into the emergent properties of the strong nuclear force~\cite{STAR:2010vob, Luo:2017faz, Bzdak:2019pkr, An:2021wof}.

Theoretical descriptions of QGP evolution in relativistic heavy-ion collisions are complex, multifaceted and computationally expensive. Relevant length scales change dynamically as the collision systems evolve. Therefore, it is necessary to match  different types of physical models in a multi-stage approach. They typically involve a combination of relativistic viscous hydrodynamics  as an effective long-wavelength description of the collective behavior and hadronic transport models~\cite{Wu_2022,Schafer:2021csj,nandi2020constraining,Schenke:2020mbo,du20203+,Nijs:2020roc,Putschke:2019yrg,Pang:2018zzo,akamatsu2018dynamically,Shen_2017, Karpenko_2015,Shen:2014vra, Petersen_2008}.   Despite the success of these models in describing collective QGP behavior, significant uncertainties persist, particularly in the initial conditions and the transport properties of the medium. These uncertainties must be quantified to extract precise information about QGP properties, such as shear and bulk viscosities~\cite{Song:2010mg, Schenke:2010rr, Qiu:2011hf, Gale:2013da, Heinz:2013th, Shen:2020mgh}.

It is challenging to compute the QGP transport coefficients from first principles~\cite{Moore:2020pfu}. There have been several attempts using lattice QCD techniques to compute the plasma's shear viscosity for a pure gluon system~\cite{Nakamura:2004sy, Meyer:2007ic, Astrakhantsev:2017nrs, Altenkort:2022yhb}.
On the other hand, extensive phenomenological studies were able to show that the hadronic observables measured from heavy-ion collisions are sensitive to the viscosities in the QGP~\cite{Schenke:2020mbo,Song:2010mg,Karpenko:2015xea,Shen:2015msa,Ryu:2015vwa,Schenke:2019ruo,Ryu:2017qzn}.
There have been many studies using hydrodynamics in heavy-ion collision simulations to estimate the specific shear viscosity $\eta/s$ for the QGP~\cite{Song:2010mg,Karpenko:2015xea,Niemi:2015qia,Shen:2020jwv, Shen:2020gef}.

Similarly, due to the extremely short lifetime, the initial state of heavy ion collisions is experimentally not accessible, leading to a variety of initial state models to exist \cite{Kolb:2001qz,Bartels:2002cj,Kowalski:2003hm,Lin:2004en,Hirano:2005xf, Drescher:2006pi,Drescher_2007,Petersen:2008dd,Werner:2010aa,Accardi:2012qut,Schenke:2012wb,Rybczynski:2013yba,Moreland:2014oya,vanderSchee:2015rta,Schafer:2021csj,Garcia-Montero:2023gex}.  Relativistic viscous hydrodynamics transforms the initial-state spatial inhomogeneities to anisotropies in the final-state hadrons momentum distributions. This theoretical description allows us to access information about the initial conditions of heavy-ion collisions and transport properties of the QGP from the experimental measurements. As many initial state models are motivated by physics but still rich in tunable parameters, and as final state anisotropies are both sensitive to the initial state inhomogeneities and to the viscosities, a large parameter space in the initial space might allow for fitting observables while setting the parameters for the initial state to values without physical motivation.

An example for an approach using an initial state with a minimal set of parameters is SMASH-vHLLE-hybrid, which uses hadronic transport both for the late stage rescattering, but also to generate the initial energy and charge density \cite{Schafer:2021csj}. There is no Bayesian analysis available for yet which uses this initial state. Therefore, such a study extends our understanding of theoretical uncertainty in the results obtained from Bayesian analysis. SMASH-vHLLE-hybrid has access to the full (3+1)D evolution and information about initial state transverse momentum, which has been shown to be relevant especially for the flow of central collisions \cite{Gotz:2023kkm}. Additionally, due to conservation of all charges and the successful reproduction of experimental results across a wide range of collision energies, it allows for the study of temperature and baryochemical dependent viscosities \cite{Gotz:2022naz}.

Recent efforts to determine the QGP's transport coefficients have focused on Bayesian inference techniques, which offer a robust framework for constraining these properties by integrating experimental data with theoretical models~\cite{Pratt:2015zsa, Bernhard:2016tnd, Auvinen:2017fjw, Bernhard:2019bmu, Nijs:2020ors, JETSCAPE:2020shq, JETSCAPE:2020mzn, Parkkila:2021yha, Parkkila:2021tqq, Phillips:2020dmw, Heffernan:2023gye, Heffernan:2023utr}. These studies have significantly advanced our understanding of QGP properties, particularly by incorporating the temperature and baryon chemical potential dependence of the specific shear and bulk viscosities into high-dimensional model parameter spaces. Nevertheless, the uncertainties remain substantial.

To leverage the extensive data from RHIC BES, we conduct comprehensive modeling of (3+1)D QGP dynamics within a 14-dimensional model parameter space, utilizing state-of-the-art relativistic viscous hydrodynamics and hadronic transport simulations. This is both a considerably smaller parameter space than current efforts \cite{jahan2024bayesiananalysis31drelativistic,shen2023viscositiesbaryonrichquarkgluonplasma} as well as larger than earlier works, such as Ref.~\cite{Auvinen:2017fjw}, which explored a much smaller five-dimensional parameter space. Due to the considerably decreased number of parameters for the initial condition, we can instead focus on a more complete study of the functional form of viscosities.

In this work, we build upon these developments by performing a systematic Bayesian inference analysis of RHIC BES measurements using full (3+1)D event-by-event simulations. Following the insights into the performance of different emulation strategies, we employ state-of-the-art Gaussian Process model emulators, enhancing the precision of our predictions~\cite{Roch:2024xhh}. This analysis will report and analyze the posterior distribution for all model parameters, providing a thorough investigation of the QGP's transport coefficients.  The results will include a detailed sensitivity analysis between model parameters and experimental observables, shedding light on the intricate dependencies that govern QGP behavior in relativistic heavy-ion collisions.

This work is structured as follows: In  \cref{sec:model}, the hybrid approach {SMASH-vHLLE-hybrid}, within which this study is performed, is briefly summarized. \Cref{sec:setup} demonstrates the choice of the parameterisation of the viscosities, as well as the choice of systems and observables. In  \cref{sec:method}, an outline is given of the setup of the Bayesian analysis employed in this study. Lastly, \cref{sec:results} demonstrates the sensitivity of different parameters to experimental observables, and shows the posterior distribution of optimal parameters, depending on the set of experimental data. From the MAP predictions, we validate that we can successfully reproduce experimental data. To conclude, a brief summary and outlook can be found in \cref{sec:Conclusion}.

\section{Model Description}\label{sec:model}

The theoretical calculations presented in this work utilize the {SMASH-vHLLE-hybrid} approach \cite{hybridurl}, a publicly available framework designed for the theoretical study of heavy-ion collisions within the energy range of $\sqrt{s_{NN}} = 4.3$ GeV to $\sqrt{s_{NN}} = 5.02$ TeV. The {SMASH-vHLLE-hybrid} model has demonstrated robust agreement with experimental data across a broad spectrum of collision energies while ensuring the conservation of all conserved charges ($B, Q, S$). It is particularly effective in capturing longitudinal baryon dynamics at intermediate collision energies \cite{Schafer:2021csj}, which motivated its development.

A concise overview of the model's components is provided below. For more detailed descriptions, readers are referred to \cite{Schafer:2021csj}.

\subsection{Initial Conditions}

The input to the fluid-dynamic evolution is the initial condition, which characterizes the system's evolution prior to the onset of hydrodynamic applicability. SMASH generates these initial conditions by simulating the interactions of particles from the colliding nuclei until each particle reaches a fixed proper time $\tau_0$. At this point, the particle is removed from the simulation. This process continues until all particles reach a hypersurface of constant proper time $\tau_0$. The transition from Cartesian coordinates (used by SMASH) to Milne coordinates (used for hydrodynamic evolution) necessitates this procedure. In this work, we investigate the optimal value for $\tau_0$ proportional to the passing time of the nuclei, $\tau_{0,\text{pass}}$: \begin{equation} \tau_{0,\text{pass}} = \frac{R_P + R_T}{\sqrt{\left(\frac{\sqrt{s_{NN}}}{2 m_N}\right)^2-1}}, \end{equation} where $R_P$ and $R_T$ are the radii of the projectile and target nuclei, respectively, $\sqrt{s_{NN}}$ is the collision energy, and $m_N$ is the nucleon mass. $\tau_{0,\text{pass}}$ is  an estimate for the earliest time when equilibration can happen, as at this time, all nucleii had the chance to interact.

To initialize the hydrodynamic evolution, the energy-momentum and charge depositions from the particles crossing the iso-$\tau$ hypersurface are smoothed to prevent shock waves. A Gaussian smearing kernel \cite{Karpenko_2015} is applied, with parameters obtained from \cite{Schafer:2021csj}. These parameters are critical for tuning the initial conditions to match experimental data, as demonstrated in similar studies \cite{Auvinen:2013sba}. 

\subsection{Hydrodynamic Evolution}

The 3+1D viscous hydrodynamics code {vHLLE} \cite{Karpenko_2014} is used to simulate the evolution of the hot and dense fireball. It solves the following hydrodynamic equations: \begin{equation} \partial_\nu T^{\mu \nu} =0, \end{equation} \begin{equation} \partial_\nu j^\nu_B=0, \quad \partial_\nu j_Q^\nu=0, \quad \partial_\nu j_S^\nu=0, \end{equation} which describe the conservation of energy, momentum, net-baryon number, net-charge, and net-strangeness. The energy-momentum tensor is decomposed as: \begin{equation} T^{\mu\nu} = \epsilon u^\mu u^\nu -\Delta^{\mu \nu}(p+\Pi) + \pi^{\mu\nu}, \end{equation} where $\epsilon$ is the local rest frame energy density, $p$ is the equilibrium pressure, $\Pi$ is the bulk pressure, and $\pi^{\mu\nu}$ represents the shear stress tensor. These equations are solved using the second-order Israel-Stewart formalism \cite{Denicol_2014, Ryu_2015}, including only second order coefficients for bulk and shear stress.

At this stage, particles are converted into fluid elements, which evolve based on a chiral mean-field equation of state \cite{Steinheimer:2011ea, Motornenko:2019arp, Most:2022wgo}, matched to a hadron resonance gas at lower energy densities. While derived from the parity doublet model, the equation of state aligns well with lattice QCD predictions and constraints from astrophysics. Evolution continues until the energy density reaches a switching value, a technical parameter optimized in this work. The freeze-out hypersurface is then constructed using the {CORNELIUS} subroutine \cite{Huovinen_2012}, and thermodynamic properties are calculated using the {SMASH} hadron resonance gas equation of state \cite{Schafer:2021csj} to ensure continuity of energy-momentum flow through the hypersurface.

\subsection{Particle Sampler}

To transition from fluid elements to particles, the {SMASH-vHLLE} hybrid approach employs the {SMASH-hadron-sampler} \cite{samplerurl}. This sampler uses the grand-canonical ensemble to particlize each surface element independently. Hadrons are sampled based on a Poisson distribution, with the mean set to the thermal multiplicity. Momenta are sampled according to the Cooper-Frye formula \cite{cooper1975landau}. Corrections to the distribution function $\delta f_{\mathrm{visc}}$ due to finite shear and bulk viscosity are incorporated using the Grad's 14-moment ansatz, assuming uniform corrections across all hadron species and neglecting charge diffusion. These corrections depend implicitly on the temperature $(T)$ and baryochemical potential $(\mu_B)$. The result is a particle list compatible with {SMASH}, enabling subsequent evolution. A comprehensive description of the sampling procedure is available in \cite{Karpenko_2015}.

\subsection{Hadronic Transport}

The {SMASH} model \cite{Weil_2016, dmytro_oliinychenko_2021_5796168, smashurl} solves the relativistic Boltzmann equation by modeling the collision integral through the formation, decay, and scattering of hadronic resonances. The model includes all hadrons listed by the Particle Data Group (PDG) up to a mass of 2.35 GeV \cite{ParticleDataGroup:2018ovx}. Initial nucleon-nucleon, as well as other hard scatterings are handled by {Pythia 8} \cite{Sj_strand_2008}, while soft interactions use a string model. For these calculations, the SMASH version 3.1 was utilized.

As the system expands, cools down and converts to hadronic phase, the Boltzmann equation with hadronic DOF becomes applicable. At the same time, as the mean free path becomes longer, the applicability of medium (fluid) picture gradually ceases.  SMASH handles hadronic rescatterings and resonance decays that happen after the particlization hypersurface. In order to do that, it propagates the input hadrons back to a common starting time and assigns appropriate formation times so that the hadrons are only allowed to interact or decay beyond the particlization surface. As the hadronic system expands further, the mean distance between the hadrons grows such that the interactions gradually cease.

In the fluid approximation, viscosities of the medium are essentially input parameters to the modelling. Different from that, in the hadronic phase, the viscosities are determined by the cross sections, which we consider as fixed. Therefore, it is inaccessible to the Bayesian Analysis which follows.

\section{Setup}
\label{sec:setup}

This section outlines the essential prerequisites for conducting a Bayesian analysis, focusing on the selection of input parameter ranges and their parameterization. The setup ensures a comprehensive exploration of the model's predictive capabilities across key aspects of the QCD phase diagram.

\subsection{Priors and Parameterization}

We consider prior distributions for three distinct parameter groups: (1) technical parameters defining the interfaces between the different stages of the hybrid approach, (2) parameters for shear viscosity, and (3) parameters for bulk viscosity. Each parameter set is discussed in detail below.

\subsubsection{Technical Parameters}

The technical parameters govern the interaction between the initial hadronic transport stage and the subsequent hydrodynamic evolution. These include two smearing parameters: $R_\perp$, for transverse smearing, and $R_\eta$, for longitudinal smearing. For the original definition of SMASH-vHLLE-Hybrid, these parameters were energy-dependent to match experimental data. However, in this study, we absorb all collision energy dependence into the temperature and baryochemical potential dependencies of viscosities. As such, $R_\perp$ and $R_\eta$ are treated as constants within the range [0.2, 2.2] fm and [0.2, 3.0], respectively. The lower limit reflects finite grid resolution constraints, while the upper limit avoids excessive smoothing, which could erase essential fluctuations.

The initialization time, $\tau_0$, is chosen proportional to the passing time of the nuclei. We assume that hydrodynamic evolution should begin only after the nuclei have passed in order for all participant scatterings to contribute to the medium production. However, $\tau_0$ must remain significantly shorter than the system's evolution time to capture relevant dynamics. Accordingly, the proportionality factor is selected within the range [0.8, 2.5]. At 200 GeV, this leads to an initialization time in the range of 0.4 fm - 1.25 fm, whereas at 7.7 GeV, the range is 2.64 fm - 8.25 fm.

The switching energy density, $\epsilon_{\text{switch}}$, marks the transition from hydrodynamics to transport. It must represent a regime where both descriptions are approximately valid—neither too dilute for hydrodynamics nor too dense for transport. This parameter is constrained to [0.25, 0.75] $\frac{\text{GeV}}{\text{fm}^3}$. 

Additionally, SMASH has a parameter which scales all cross-sections, which we call $\sigma_{\rm AB, \rm{scale}}$. Although the cross-sections in SMASH are tuned with respect to experimental data, this allows to test whether in a hybrid approach, different values of the cross-sections are preferred.

~\Cref{tab:parameters} summarizes the prior ranges for all technical parameters.

\begin{table}[h!]
    \centering
    \begin{tabular}{c|c c}
        \hline \hline
        \textbf{Parameter} & \textbf{Prior Range} &\\ \hline
        $R_{\perp}$           & [0.2, 2.2] & fm \\ 
        $R_{\eta}$            & [0.2, 3.0] & fm \\ 
        $\tau_{\text{IC,scale}}$              & [0.8, 2.5] &  \\ 
        $\epsilon_{\text{switch}}$ & [0.25, 0.75] & $\frac{\text{GeV}}{\text{fm}^3}$ \\ 
        $\sigma_{\rm AB, \rm{scale}}$& [0.8, 1.2] & \\\hline \hline
    \end{tabular}
     \caption{Priors for technical parameters.}
     \label{tab:parameters}
\end{table}

\subsubsection{Shear Viscosity}

The shear viscosity parameterization accounts for temperature and baryochemical potential dependence:
\begin{align}  \nonumber 
    \frac{\eta T}{\epsilon + P} =&   \max \left( 0,(\eta/s)_{\text{kink}}  +\label{eq:shear} \begin{cases}
a_{l,\eta}(T - T_c) & T < T_c,\\
a_{h,\eta}(T - T_c) & T > T_c \\
\end{cases}\right)\\   
& \times \left(1 + a_{\mu_B}\frac{\mu_B}{\mu_{B,0}}\right),\\   
T_c &= T_{\eta,0} + b_{\mu_B} \frac{\mu_B}{\mu_{B,0}}.
\end{align}

\begin{figure}
    \centering
    \includegraphics[width=\linewidth]{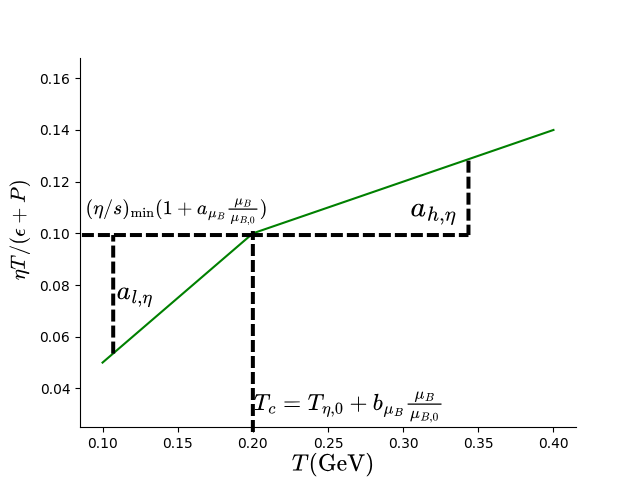}
    \caption{Illustration of the shear viscosity parameterization.}
    \label{fig:shear_explanation}
\end{figure}

$\mu_{B,0}$ is here a normalization factor in the form of the baryochemical potential of cold nuclear matter. The parameters are illustrated in ~\cref{fig:shear_explanation}. The parameter $(\eta/s)_{\text{min}}$ sets the shear viscosity at $T_c$ for $\mu_B = 0$, with $a_{l,\eta}$ and $a_{h,\eta}$ controlling the slopes in the low- and high-temperature regions, respectively. The critical temperature $T_c$ varies with $\mu_B$, governed by $T_{\eta,0}$ and $b_{\mu_B}$. The overall viscosity is scaled by $a_{\mu_B}$ to reflect $\mu_B$-dependent effects.

This flexible structure accommodates a temperature dependence consistent with earlier studies \cite{JETSCAPE:2020shq, JETSCAPE:2020mzn}, while explicitly incorporating $\mu_B$ effects.  We allow for an explicit temperature dependence, greatly increasing the flexibility with respect to \cite{shen2023viscositiesbaryonrichquarkgluonplasma} which featured only mild, linear temperature dependence. This allows us to test whether earlier agreements with constant, temperature independent shear viscosity holds in our model as well \cite{JETSCAPE:2020mzn}. 

The prior ranges (\cref{tab:parameters_shear}) are broad enough to include both $\mu_B$-independent and constant shear viscosities. The KSS-bound of 0.08 is not enforced, allowing the analysis to explore the parameter space freely.

\begin{table}[h!]
    \centering
    \begin{tabular}{c|c c}
        \hline \hline
        \textbf{Parameter} & \textbf{Prior Range} & \\ \hline
        $a_{l,\eta}$           & [-15, 1] & $\frac{1}{\text{GeV}}$\\ 
        $a_{h,\eta}$           & [-15, 1.75] & $\frac{1}{\text{GeV}}$\\ 
        $T_{\eta,0}$           & [0.09, 0.25] & GeV  \\ 
        $(\eta/s)_{\text{min}}$ & [0.001, 0.35] & \\ 
        $a_{\mu_B}$     & [-0.8, 7] & \\ 
        $b_{\mu_B}$     & [-0.3, 0.8] & \\ \hline \hline
    \end{tabular}
    \caption{Priors for shear viscosity parameters.}
    \label{tab:parameters_shear}
\end{table}

\subsubsection{Bulk Viscosity}

\begin{figure}
    \centering
    \includegraphics[width=\linewidth]{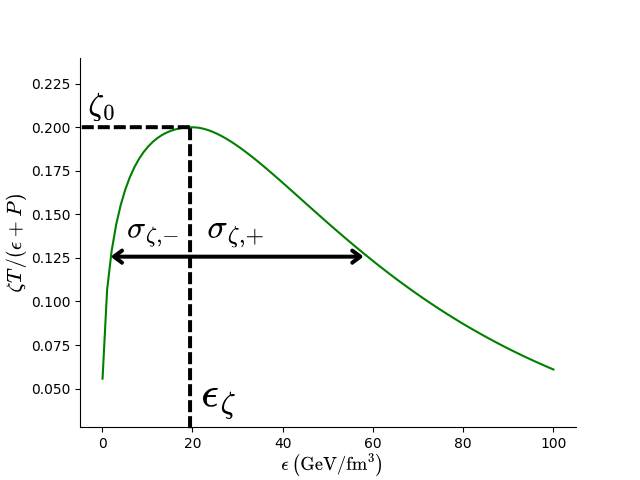}
    \caption{Illustration of the bulk viscosity parameterization.}
    \label{fig:bulk_explanation}
\end{figure}

In contrast to recent studies \cite{shen2023viscositiesbaryonrichquarkgluonplasma} we choose one bulk viscosity parameterization for all collision energies. The bulk viscosity parameterization incorporates $\mu_B$ implicitly by depending on the energy density $\epsilon$. This approach assumes that the bulk viscosity peak follows a line of constant $\epsilon$, potentially aligning with the QCD phase crossover at finite $\mu_B$. The parameterization is:
\begin{equation}
    \frac{\zeta T}{\epsilon + P} = \zeta_0  \begin{cases}
\exp\left(-\beta\frac{(\epsilon^{1/4}-\epsilon_\zeta^{1/4})^2}{2\sigma_{\zeta,-}^2}\right), & \epsilon < \epsilon_\zeta, \\
\exp\left(-\beta\frac{(\epsilon^{1/4}-\epsilon_\zeta^{1/4})^2}{2\sigma_{\zeta,+}^2}\right), & \epsilon > \epsilon_\zeta,
\end{cases} 
\end{equation}
where $\zeta_0$ controls the peak amplitude, $\epsilon_\zeta$ sets the location of the peak, and $\sigma_{\zeta,-}$ and $\sigma_{\zeta,+}$ define the widths below and above $\epsilon_\zeta$, respectively. As we want to have a similiar shape than comparable parameterization in the temperature, we take the fourth root of the energy density and scale it by a conversion factor, which is called $\beta$. \Cref{fig:bulk_explanation} illustrates this parameterization, while \cref{tab:parameters_bulk} lists the prior ranges.

\begin{table}[h!]
    \centering
    \begin{tabular}{c|cc}
        \hline \hline
        \textbf{Parameter} & \textbf{Prior Range} \\ \hline
        $\zeta_0$           & [0, 0.2] & \\ 
        $\epsilon_\zeta$    & [0.5, 40] & $\frac{\text{GeV}}{\text{fm}^3}$ \\ 
        $\sigma_{\zeta,-}$  & [0.005, 0.1] & \\ 
        $\sigma_{\zeta,+}$  & [0.01, 0.15] & \\ \hline \hline
    \end{tabular}
    \caption{Priors for bulk viscosity parameters.}
    \label{tab:parameters_bulk}
\end{table}

\subsection{Systems}

We investigate three collision energies—7.7 GeV, 19.6 GeV, and 200 GeV—spanning a broad region of the QCD phase diagram. To optimize computational efficiency, we limit the analysis to three centrality classes: 0–5\% (central), 15–25\% (only for rapditiy-dependent data at 200 GeV) , and 20–30\% (mid-central). This selection captures key dynamics while reducing computational costs. The underlying assumption is that a model which successfully describes two centrality classes is also able to describe all other classes.

These systems provide a wide $\mu_B$ coverage of mean values of around 0 to 400 MeV at freeze-out \cite{Cleymans:2005xv, Andronic:2009jd, STAR:2017sal}, offering insights into $\mu_B$-dependent effects. Observables sensitive to longitudinal dynamics further enhance this range \cite{Neff:2024jrd}.
\subsection{Observables}
\Cref{table:expData} summarizes the experimental observables. 
They consist of STAR RHIC BES data integrated bulk, enriched with $dN/d\eta$ from PHOBOS at $\sqrt{s_{NN}} = $ 200 GeV. For 19.6 GeV, there is a choice between $dN/d\eta$ data from PHOBOS and STAR, which show substantial differences \cite{Molnar:2023htb}. We chose the more recent STAR data for our tuning. $\eta$-differential data is included until a cutoff of 3 in forward and backwards-rapidity. The total dataset size is 168 datapoints.

For the identified particle observables, we have excluded antiprotons due to the high statistical uncertainties. Additionally, the initialisation with hadronic transport leads to excess baryon charge at 200 GeV, due to which we have excluded protons at 200 GeV \cite{Garcia-Montero:2021haa}.

\begin{table*}[t]
 \centering
    \begin{tabular}{c|c|c | c} \hline \hline
        $\sqrt{s_{NN}}$ & 0-5\% & 15-25\%  & 20-30\%  \\ \hline
             & $dN/dy|_{y=0} (\pi^{+,-}, K^{+,-})$ \cite{STAR:2008med}& & $dN/dy|_{y=0} (\pi^{+,-}, K^{+,-})$ \cite{STAR:2008med} \\
             200 GeV & $\langle p_T \rangle|_{y=0} (\pi^{+,-}, K^{+,-})$ \cite{STAR:2008med} &  & $\langle p_T \rangle|_{y=0} (\pi^{+,-}, K^{+,-})$ \cite{STAR:2008med} \\
             & $v^\mathrm{ch}_2\{2\}|_{y=0}$~\cite{STAR:2017idk}, $v^\mathrm{ch}_3\{2\}|_{y=0}$~\cite{STAR:2016vqt} & & $v^\mathrm{ch}_2\{2\}|_{y=0}$~\cite{STAR:2017idk}, $v^\mathrm{ch}_3\{2\}|_{y=0}$~\cite{STAR:2016vqt}\\
            & $dN^\mathrm{ch}/d\eta$ \cite{PHOBOS:2006wwo}, $v_2^\mathrm{ch}(\eta)$ \cite{PHOBOS:2004vcu} & $dN^\mathrm{ch}/d\eta$ \cite{PHOBOS:2006wwo}, $v_2^\mathrm{ch}(\eta) $\cite{PHOBOS:2004vcu} & \\
            \hline 
             & $dN/dy|_{y=0} (\pi^{+,-}, K^{+,-}, p)$ \cite{STAR:2017sal}&  & $dN/dy|_{y=0} (\pi^{+,-}, K^{+,-}, p)$ \cite{STAR:2017sal} \\
          19.6 GeV   & $\langle p_T \rangle |_{y=0}(\pi^{+,-}, K^{+,-}, p)$ \cite{STAR:2017sal} & & $\langle p_T \rangle |_{y=0}(\pi^{+,-}, K^{+,-}, p)$ \cite{STAR:2017sal} \\
           & $dN^\mathrm{ch}/d\eta$ \cite{Molnar:2023htb}&  & $dN^\mathrm{ch}/d\eta$ \cite{Molnar:2023htb}\\
           & $v^\mathrm{ch}_2\{2\}|_{y=0}$~\cite{STAR:2017idk}, $v^\mathrm{ch}_3\{2\}|_{y=0}$~\cite{STAR:2016vqt} & & $v^\mathrm{ch}_2\{2\}|_{y=0}$~\cite{STAR:2017idk}, $v^\mathrm{ch}_3\{2\}|_{y=0}$~\cite{STAR:2016vqt} \\ \hline
             & $dN/dy|_{y=0} (\pi^{+,-}, K^{+,-}, p)$ \cite{STAR:2017sal}& & $dN/dy|_{y=0} (\pi^{+,-}, K^{+,-}, p)$ \cite{STAR:2017sal} \\ 
        7.7 GeV & $\langle p_T \rangle|_{y=0} (\pi^{+,-}, K^{+,-}, p)$ \cite{STAR:2017sal}& &  $\langle p_T \rangle|_{y=0} (\pi^{+,-}, K^{+,-}, p)$ \cite{STAR:2017sal} \\
             & $v^\mathrm{ch}_2\{2\}|_{y=0}$~\cite{STAR:2017idk}, $v^\mathrm{ch}_3\{2\}|_{y=0}$~\cite{STAR:2016vqt} & & $v^\mathrm{ch}_2\{2\}|_{y=0}$~\cite{STAR:2017idk}, $v^\mathrm{ch}_3\{2\}|_{y=0}$~\cite{STAR:2016vqt} \\
        \hline \hline     
    \end{tabular}
    \caption{The experimental measurements in Au+Au collisions used in this Bayesian inference study.}
    \label{table:expData}
\end{table*}
\section{Method}
\label{sec:method}

This section describes the methodology employed to perform Bayesian parameter inference within the SMASH-vHLLE-hybrid framework, encompassing parameter sampling, emulation, and posterior distribution estimation.

\subsection{Parameter Sampling and Event Generation}

To explore the 15-dimensional parameter space, we employ Latin Hypercube Sampling (LHS), implemented via pyDOE \cite{pydoeurl,DEUTSCH2012763} and included in SMASH-vHLLE-Hybrid  \cite{hybridurl}. This method ensures an efficient and uniform coverage of the prior space. A total of 750 parameter points are sampled, and for each, we generate 250 events per centrality class.

The centrality classes are selected by generating 5000 initial state events, using the energy and baryon number distribution to calculate the entropy, and selecting events accordingly. It is important to note that this is only an approximation of the centrality classes used in experiment. To achieve equal statistics for all centrality classes, only half of the events are chosen for the wider classes. These initial events are propagated through all stages of the hybrid approach. To improve statistical accuracy and maintain charge conservation, oversampling is performed at different magnitudes depending on the collision energy: 4000, 2000, and 500 events for $\sqrt{s_{NN}} = 7.7$, $19.6$, and $200$ GeV, respectively.

\subsection{Observable Analysis}

Analysis of the generated events is conducted using the SPARKX Python package in version 2.0.2 \cite{Sass:2025opk,hendrik_roch_2024_12821236}, which automates the computation of bulk observables and flow coefficients. This reliable and efficient framework facilitates a direct comparison between model predictions and experimental data.

\begin{figure*}
    \centering
    \includegraphics[width=\linewidth]{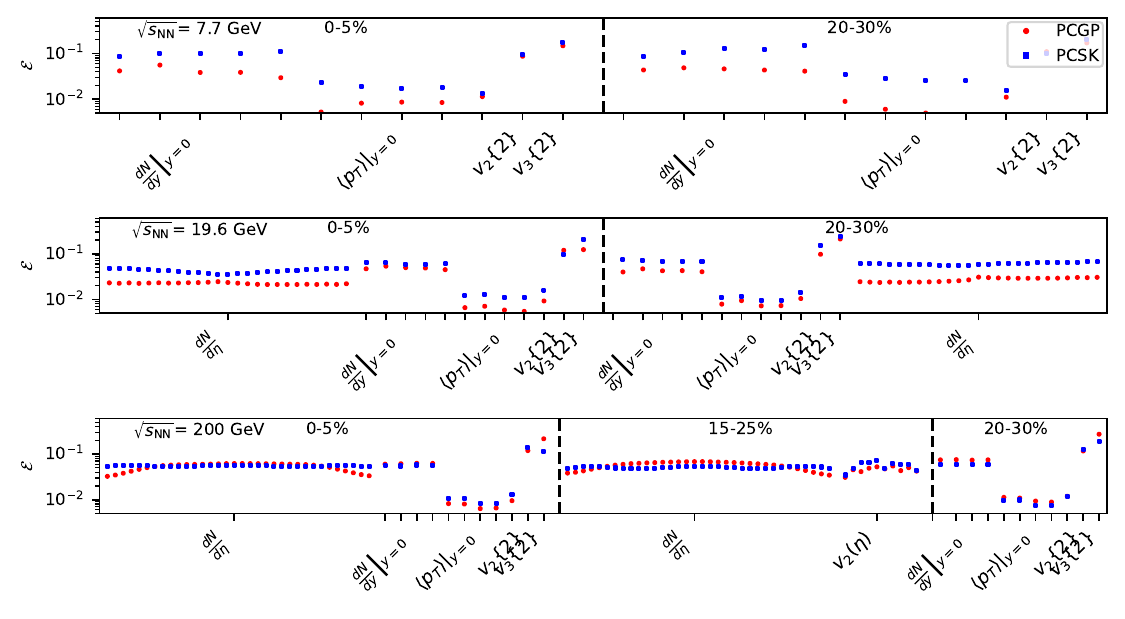}
    \caption{The averaged root mean square error $\mathcal{E}$ for both emulation strategies for all observables.}
    \label{fig:emu_validation_rms}
\end{figure*}

\begin{figure*}
    \centering
    \includegraphics[width=\linewidth]{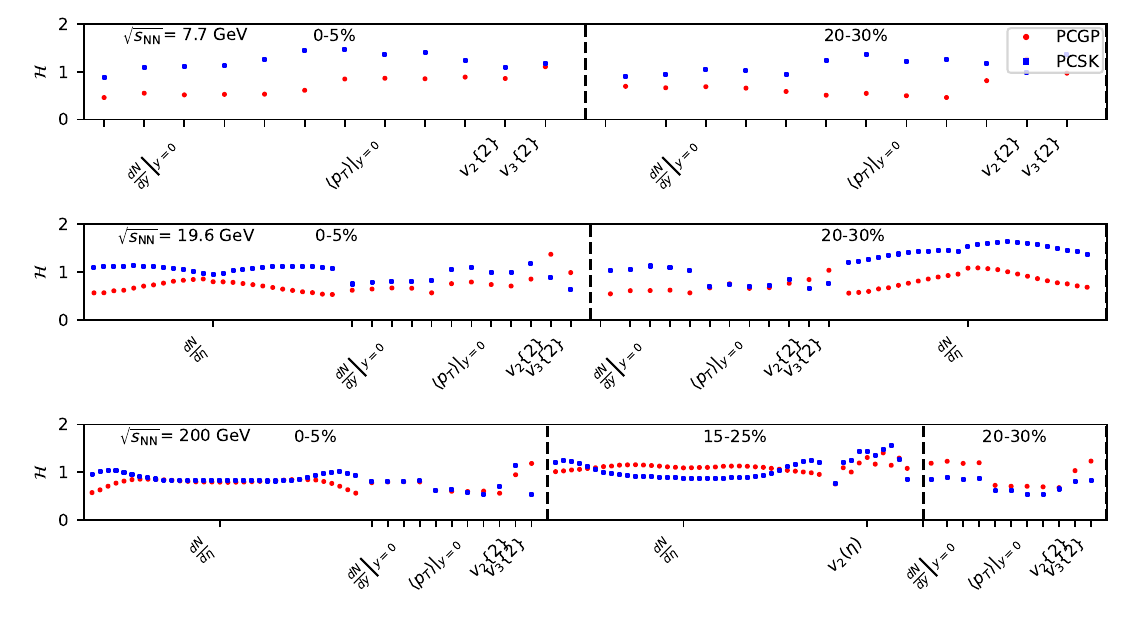}
    \caption{The uncertainty estimation metric $\mathcal{H}$ for both emulation strategies for all observables.}
    \label{fig:emu_validation_hon}
\end{figure*}

\begin{figure*}
    \centering
    \includegraphics[width=\linewidth]{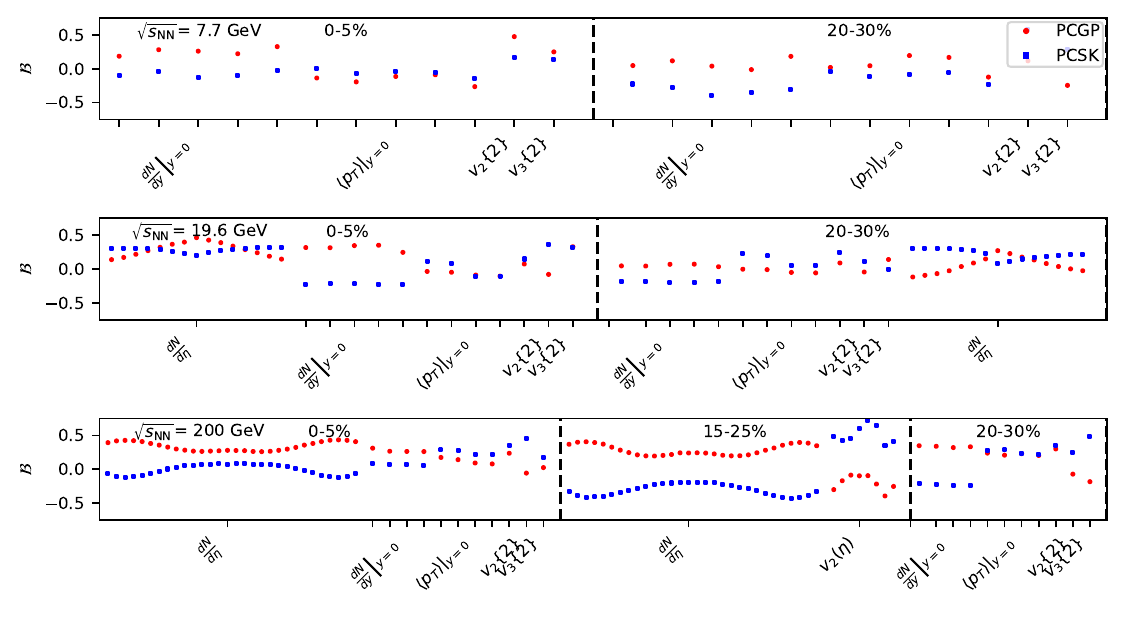}
    \caption{The bias $\mathcal{B}$ for both emulation strategies for all observables.}
    \label{fig:emu_validation_bias}
\end{figure*}

\begin{figure*}
    \centering
    \includegraphics[width=\linewidth]{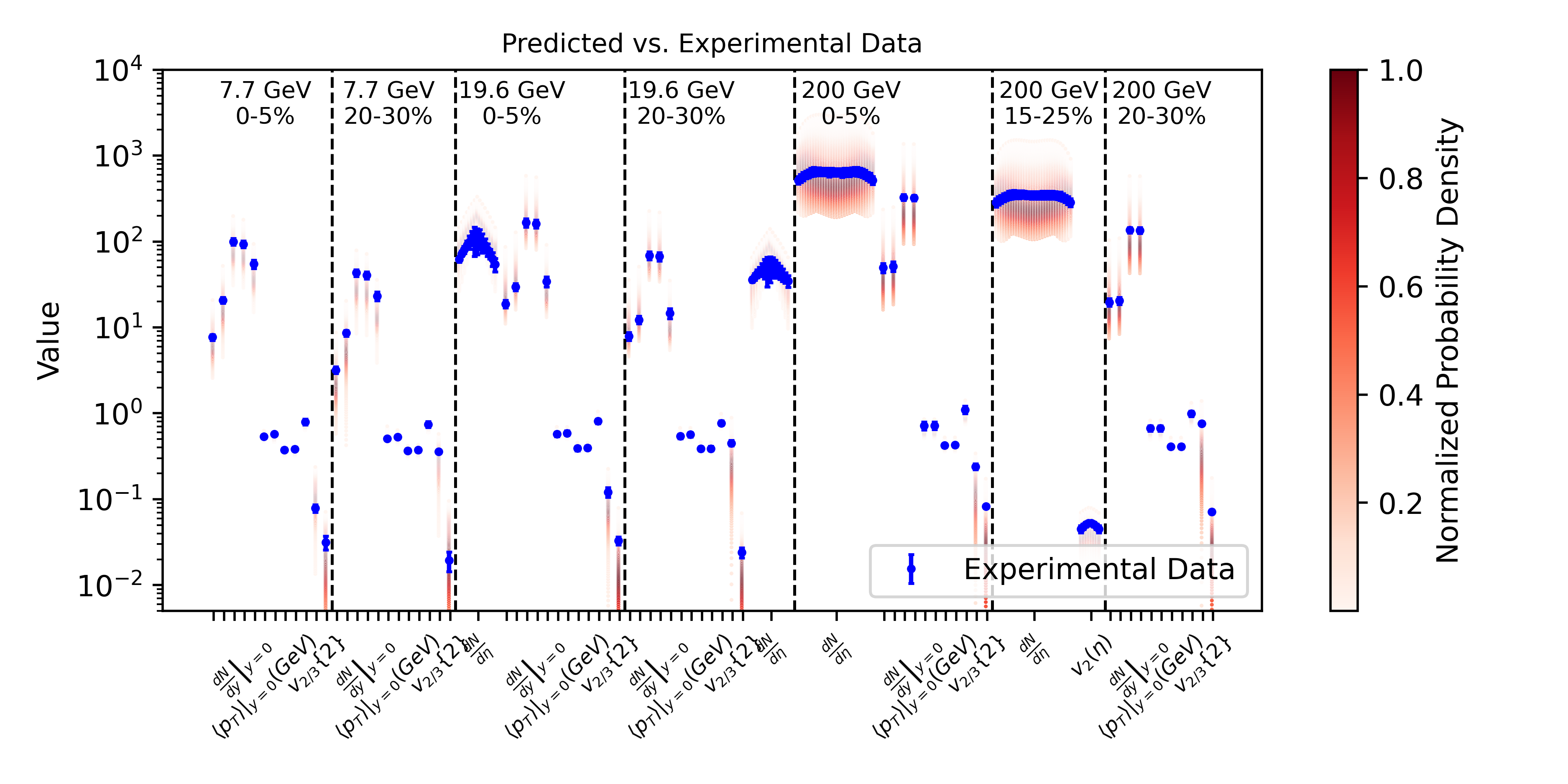}
    \caption{Probability distribution of the posterior observables for the full prior range, compared to the experimental data points used in this study. For mean transverse momenta, the width of the distribution is comparable to the size of the data markers. }
    \label{fig:priordistri}
\end{figure*}

\subsection{Bayesian Framework}

The primary goal of the Bayesian analysis is to identify the optimal parameter set within the prior range such that the model predictions closely match experimental data. Let $\bs{\theta} \equiv (\theta_1, \dots, \theta_m)$ represent the model parameters, and let ${\bf y}_{\mathrm{sim}}(\bs{\theta})$ denote the simulation output in $\mathbb{R}^d$, where $d$ is the dimension of the observable space. These outputs are compared to the experimental values ${\bf y}_{\mathrm{exp}} \equiv (y_{\mathrm{exp}, 1}, \dots, y_{{\mathrm{exp}}, d})$, forming a statistical model:
\begin{equation}
{\bf y}_{\mathrm{exp} } = {\bf y}_{\mathrm{ sim}}(\bs{\theta}) + \bs{\epsilon},
\label{eq:model}
\end{equation}
where $\bs{\epsilon}$ is the residual error, assumed to follow a multivariate normal (MVN) distribution with mean $\mathbf{0}$ and covariance $\bs{\Sigma}$, which is the experimental covariance.

The Bayesian framework updates prior beliefs about the parameters $\bs{\theta}$ using Bayes' theorem:
\begin{equation}
\mathcal{P}(\bs{\theta} | {\bf y}_{\mathrm{exp}}) = \frac{\mathcal{P}({\bf y}_{\mathrm{exp}} | \bs{\theta}) \mathcal{P}(\bs{\theta})}{\mathcal{P}({\bf y}_{\mathrm{exp}})}.
\label{eq:bayes}
\end{equation}
Here, $\mathcal{P}(\bs{\theta} | {\bf y}_{\mathrm{exp}})$ is the posterior probability density, $\mathcal{P}(\bs{\theta})$ is the prior, $\mathcal{P}({\bf y}_{\mathrm{exp}})$ is the evidence, and $\mathcal{P}({\bf y}_{\mathrm{exp}} | \bs{\theta})$ is the likelihood. Given the MVN assumption for $\bs{\epsilon}$, the likelihood is expressed as:
\begin{widetext}
\begin{equation}
\mathcal{P}({\bf y}_{\mathrm{exp}} | \bs{\theta}) = \frac{1}{\sqrt{|2\pi\bs{\Sigma}|}} \exp\left[-\frac{1}{2}({\bf y}_{\mathrm{sim}}(\bs{\theta})-{\bf y}_{\mathrm{exp}})^\mathsf{T}\bs{\Sigma}^{-1}({\bf y}_{\mathrm{sim}}(\bs{\theta})-{\bf y}_{\mathrm{exp}})\right].
\label{eq:likelihood}
\end{equation}
\end{widetext}

\subsection{Gaussian Process Emulation}

Direct evaluation of ~\cref{eq:bayes} is computationally expensive, as simulating each parameter set requires significant computational resources. We want therefore to minimise the number of parameter sets to be evaluated while at the same time minimise the loss of information.  In order to achieve this, we employ Gaussian Process (GP) emulators as surrogates for the hybrid model \cite{gramacy2020surrogates, Rasmussen2004}. These emulators are trained on the simulation outputs corresponding to the sampled parameter sets, enabling efficient predictions of mean $\bs{\mu}(\bs{\theta})$ and covariance ${\bf C}(\bs{\theta})$ for the full parameter space. 

The likelihood function in ~\cref{eq:likelihood} is approximated using the GP emulator as:
\begin{widetext}
\begin{equation}
\mathcal{P}({\bf y}_{\mathrm{exp}} | \bs{\theta}) \approx \frac{1}{\sqrt{|2\pi{\bf V}(\bs{\theta})|}} \exp\left[-\frac{1}{2}(\bs{\mu}(\bs{\theta})-{\bf y}_{\mathrm{exp}})^\mathsf{T}{\bf V}(\bs{\theta})^{-1}(\bs{\mu}(\bs{\theta})-{\bf y}_{\mathrm{exp}})\right],
\label{eq:approx_likelihood}
\end{equation}
\end{widetext}

where ${\bf V}(\bs{\theta}) = {\bf C}(\bs{\theta}) + \bs{\Sigma}$. This approach significantly reduces computational costs while maintaining high accuracy.

For this work, we utilize the BAND framework, specifically the surmise library \cite{surmise2024}.  

Using the trained GP emulators, we can obtain the posterior distribution of model parameters, $\mathcal{P}(\theta \vert y_\mathrm{exp})$, following Bayes' theorem by sampling the (in our case) uniform prior $\mathcal{P}(\theta)$ with the Monte Carlo Markov Chain (MCMC) method,
\begin{equation}
    \mathcal{P}(\theta \vert y_\mathrm{exp}) \propto \mathcal{P}(y_\mathrm{exp} \vert \theta) \mathcal{P}(\theta).
\end{equation}
Here $\mathcal{P}(y_\mathrm{exp} \vert \theta)$ is the likelihood for the model results with parameter $\theta$ to agree with the experimental data $y_\mathrm{exp}$. 

\subsection{Posterior Estimation via MCMC}

The posterior distribution is sampled using Markov Chain Monte Carlo (MCMC) methods, specifically the Preconditioned Monte Carlo (PMC) algorithm implemented in the pocoMC package \cite{karamanis2022accelerating, karamanis2022pocomc}. PMC integrates persistent sampling, normalizing flow preconditioning, and a gradient-free Markov kernel. The PMC algorithm incrementally adjusts an inverse temperature parameter, transitioning from the prior to the posterior distribution. It enables effective exploration of complex posterior distributions without relying on gradient information. This approach is particularly advantageous for high-dimensional and computationally intensive models, as it enhances sampling efficiency and accuracy.

Normalizing flow preconditioning, a powerful technique based on neural networks, allows for efficient learning of complex high-dimensional distributions. It has been successfully applied in diverse contexts, including image processing \cite{10.1145/3341156} and matrix element integration \cite{Deutschmann:2024lml}. The combination of these techniques ensures accurate exploration of the posterior distribution, even for high-dimensional and computationally intensive models.

The corresponding codebase of \cite{Roch:2024xhh} was also used for generating the emulators and performing the posterior generation.

\section{Validation}
Validation is a crucial process when performing a Bayesian analysis, as it ensures trustworthy results and spots potential issues in the parameter estimation.
\subsection{Emulator validation}
\begin{figure}
    \centering
    \includegraphics{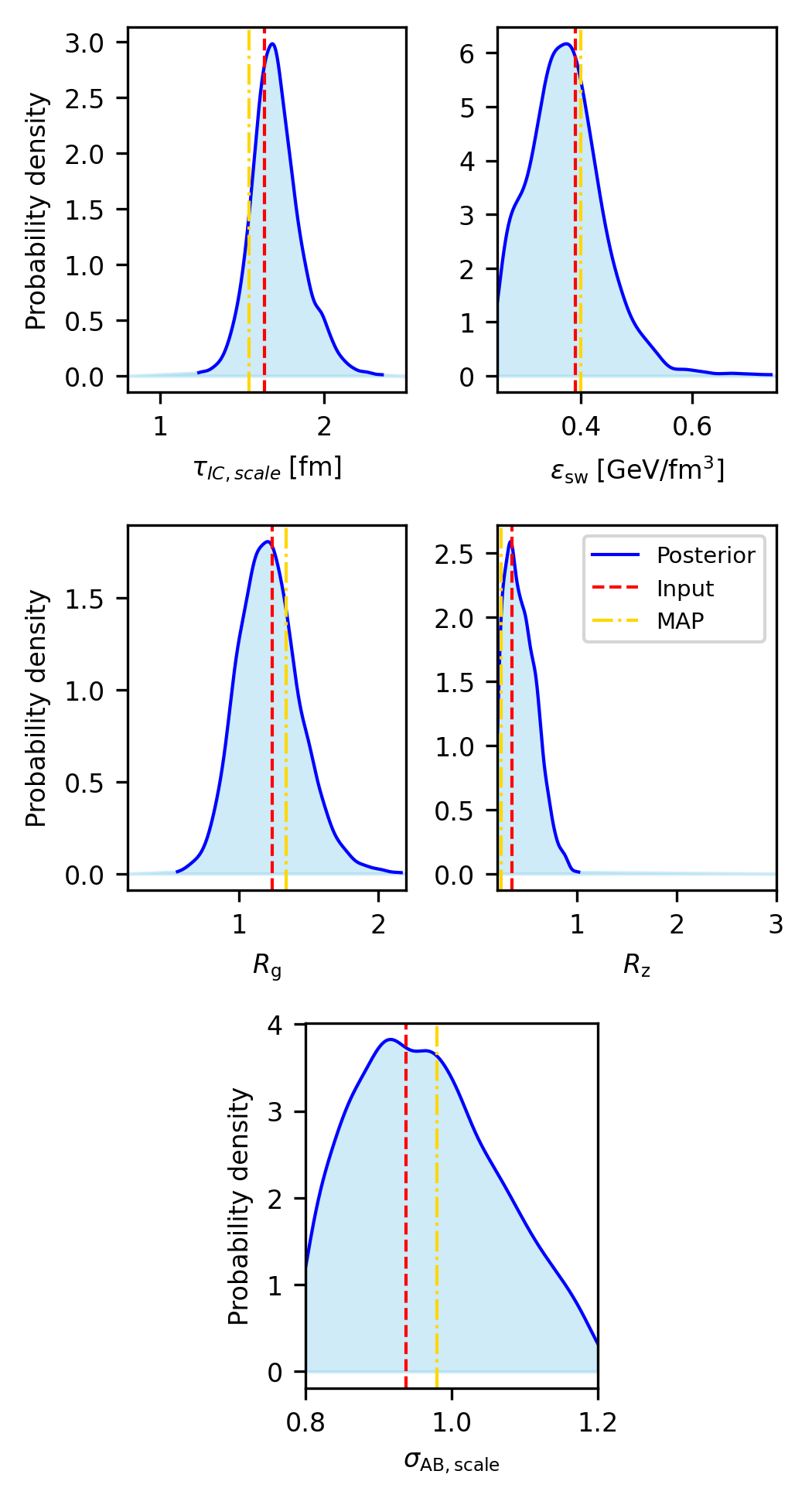}
    \caption{Posterior of the technical parameters in the closure test. The red vertical line is the value of the parameter generating the pseudo-experimental data, the golden vertical line represents the maximum-a-posteriori value.}
    \label{fig:closure_tech}
\end{figure}
\begin{figure}
    \centering
    \includegraphics{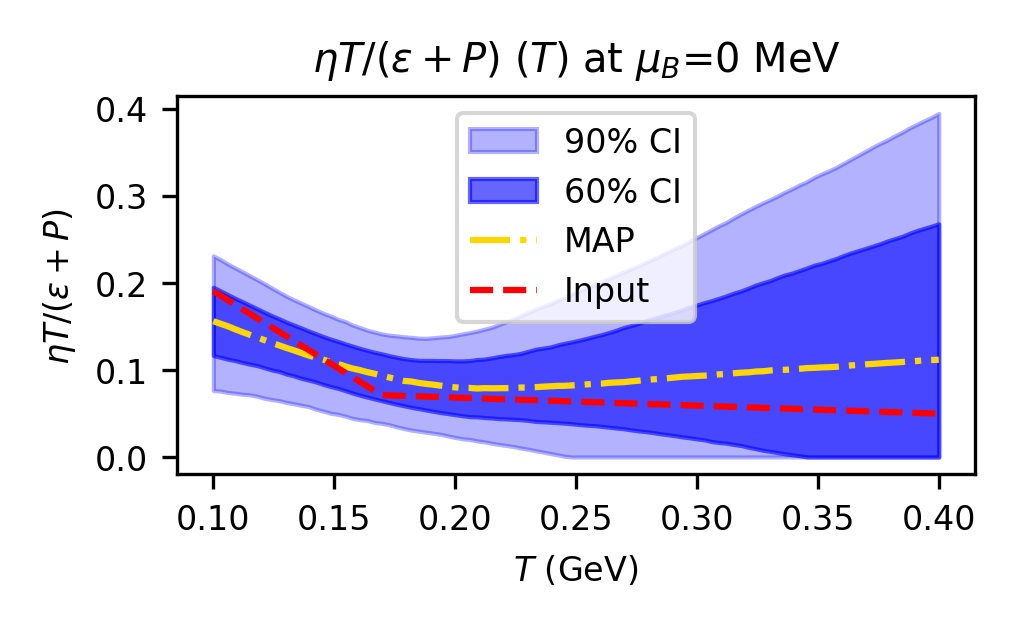}
    \includegraphics{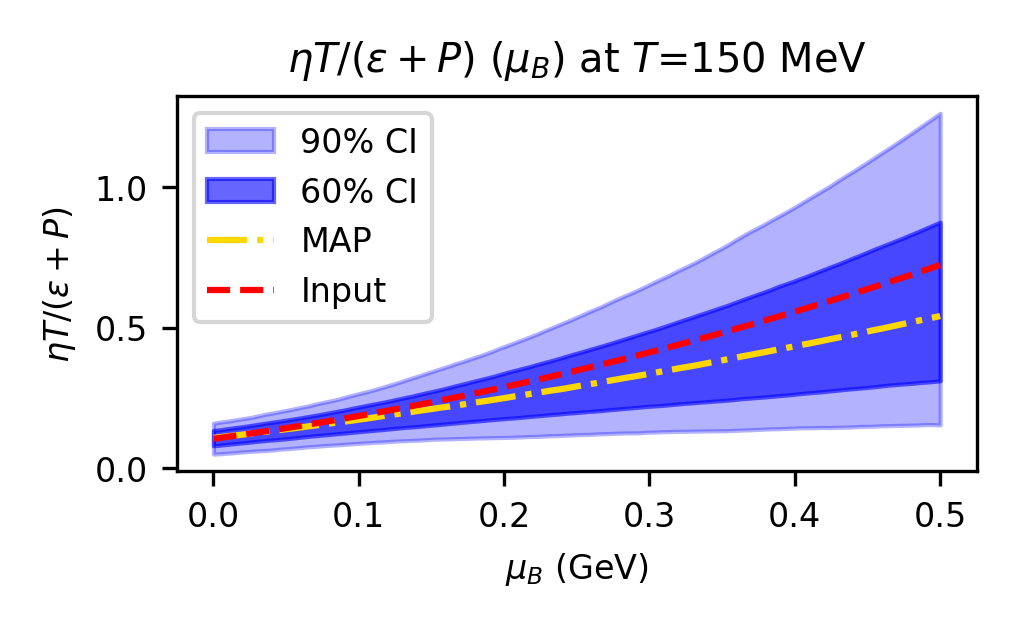}
    \includegraphics{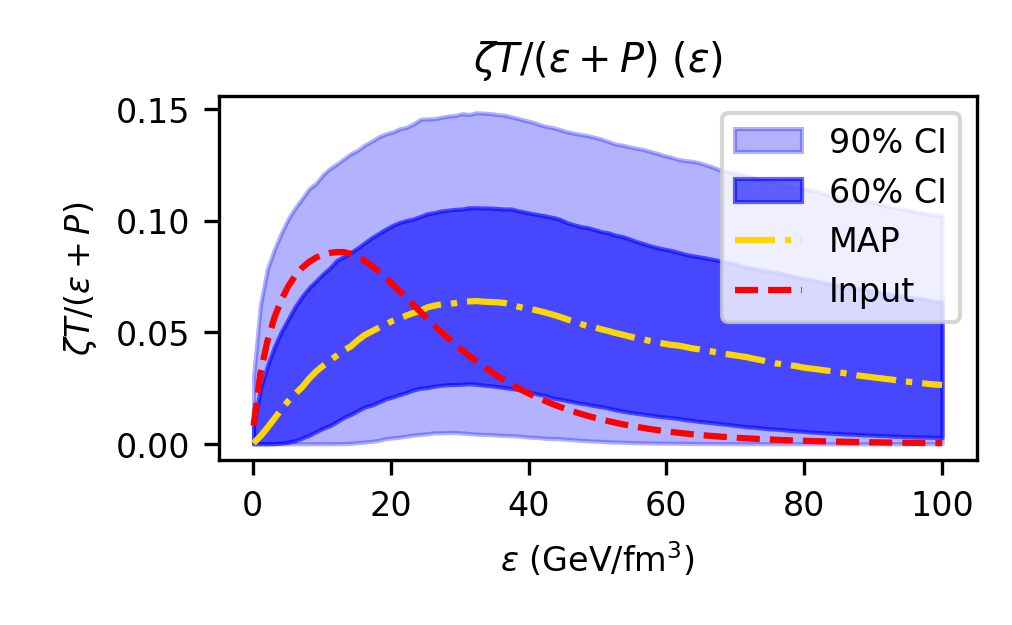}
    \caption{Posterior of the viscosities in the closure test. From top to bottom: the shear viscosity as a  function of temperature for vanishing baryochemical potential, the shear viscosity as a function of baryochemical potential at fixed temperature, and the bulk viscosity as a function of the energy density. The red line is the input parameterization, and the golden line represents the maximum-a-posteriori. The bands represent the 60\% and 90\% confidence interval of the posterior at the specified temperature/baryochemical potential.}
    \label{fig:closure_visc}
\end{figure}

We use the test metric for benchmarking proposed in \cite{Roch:2024xhh} to choose the most accurate GP configuration in our setup. 
To quantify the prediction error of the GP emulators, one defines
\begin{equation}
    \mathcal{E}\equiv\sqrt{\left\langle\left(\frac{\text{prediction}-\text{truth}}{\text{truth}}\right)^2\right\rangle}
    \label{eq:uncertainty}
\end{equation}
for each observable in the analysis, whereas the emulators uncertainty is quantified as
\begin{equation}
    \mathcal{H}\equiv\ln\left(\sqrt{\left\langle\left(\frac{\text{prediction}-\text{truth}}{\text{prediction uncertainty}}\right)^2\right\rangle}\right).
    \label{eq:honesty}
\end{equation}
Additionally, we take at the average bias in multiples of the prediction uncertainty, which quantifies whether we systematically over- or underestimate the value of an observable: 
\begin{equation}
\mathcal{B}\equiv\left\langle\left(\frac{\text{prediction}-\text{truth}}{\text{prediction uncertainty}}\right)\right\rangle.
    \label{eq:honesty}
\end{equation}
For an accurate prediction, we expect the values of $\mathcal{E}$, $\mathcal{H}$ and $\mathcal{B}$ to all approach 0.
In the case where $\mathcal{H}>0$, the emulator gives uncertainties that are too small compared to the actual error away from the true values; when $\mathcal{H}<0$, the returned uncertainty estimates are too conservative.
To determine prediction and truth, 15 parameter sets were excluded from the training of the GP. Then, the prediction for the value and its uncertainty at the excluded parameter sets are compared to the simulation results at these values.
The values of the benchmarking metric for each observable can be found in \cref{fig:emu_validation_rms}, \cref{fig:emu_validation_hon} and \cref{fig:emu_validation_bias}, for two prescriptions of GP emulation which were found in \cite{Roch:2024xhh} to perform well: PCGP (Principal Component Gaussian Process) and PCSK (Principal Component Stochastic Kriging). They both use the  PCA-decomposition and the Matérn kernel. For each principal component $t_l(\boldsymbol{\theta}) = \boldsymbol{s}_l^\mathsf{T}\,\tilde{\textbf y}_{\mathrm{sim}}(\boldsymbol{\theta})$, where $\tilde{\textbf y}_{\mathrm{sim}}$ is the standardized output, a GP model provides a predictive mean $m_l(\boldsymbol{\theta})$ and variance $s_l^2(\boldsymbol{\theta})$. In other words, each principle component is drawn from a Gaussian distribution with mean $m_l(\boldsymbol{\theta})$ and standard deviation $s_l^2(\boldsymbol{\theta})$,
where $m_l(\boldsymbol{\theta}) = \mathbf{k}_l^\mathsf{T}\mathbf{K}_l^{-1}\mathbf{t}_l$ and $s^2_l(\boldsymbol{\theta}) = k_l(\boldsymbol{\theta}, \boldsymbol{\theta}) - \mathbf{k}_l^\mathsf{T}(\boldsymbol{\theta})\mathbf{K}_l^{-1}\mathbf{k}_l(\boldsymbol{\theta})$. 
Here, $\mathbf{k}_l(\boldsymbol{\theta})=\left[k_l(\boldsymbol\theta,\boldsymbol\theta^{\mathrm tr}_i)\right]^n_{i=1}$ denotes the covariance vector between $n$ training parameters $\lbrace\boldsymbol\theta_1^{\mathrm tr},\dots,\boldsymbol\theta_n^{\mathrm tr}\rbrace$ and any chosen parameter $\boldsymbol{\theta}$, and $\mathbf{K}_l = \left[k_l(\boldsymbol\theta_i^{\mathrm tr},\boldsymbol\theta_j^{\mathrm tr}) + \delta_{i, j}r_{l, i}\right]_{i, j = 1}^n$ represents the covariance matrix between the $n$ training parameters. $r_{l,i}$ is the square of the statistical uncertainty of the $l$-th principal component at the training point $i$, and is only present for PCSK. As a result, only PCSK includes also the uncertainty of the model predictions in the emulation.

For the three panels of \cref{fig:emu_validation_rms} show, from top to bottom, the energy increases, whereas the centrality decreases from left to right, with centrality classes separated by a line. Each x-tick represents a different observable.
We observe that for both prescriptions, errors remain within acceptable bounds. The biggest deviations occur for the integrated flow, probably due to the higher impact of fluctuations. In comparison to \cite{Roch:2024xhh}, we observe basically no rapidity dependence of the errors. Additionally, for the majority of data points, PCGP outperforms PCSK substantially.
A similiar picture emerges for the uncertainty metric. It is for most datapoints considerably closer to zero than for PCSK, and therefore underestimates the uncertainty less. This is remarkable as PCSK takes the standard deviation of the training data into account, whereas PCGP does not. However, such behavior can be also observed for several observables in \cite{Roch:2024xhh}. It is currently not clear why PCSK provides less conservative estimates for the uncertainty, although it includes the uncertainty from the model. This question is left for further studies. In the meantime, we note that PCSK is not suited for giving a good uncertainty estimation.
For the bias, a less clear picture evolves. For some observables, the two different emulating strategies show a bias in different directions. In general, the bias is substantially smaller than the prediction uncertainty, underlining that the emulation process is successfull.
Of the proposed preprocessing methods, namely logscaling the observables and performing a PCA the viscosity parameterizations, none could improve the performance in these metrics. Therefore, they were not employed.
As PCGP outperforms PCSK in most metrics, we continue using this approach for the rest of this work.

\subsection{Prior validation}
A further validation is the consideration if the prior range chosen is sufficient to fit experimental data. \Cref{fig:priordistri} shows the probability distribution for values of the observables to be fitted to experimental data. In simple words, it shows the likelihood for each observable to be predicted if a random point were to be chosen from prior space. We see that the experimental data points lie all within the this range. However, we see that certain experimental datapoints lie closer to the edges of the range, especially flows at  higher collision energies  and $\frac{\mathrm{d}N}{\mathrm{d}\eta}$ at 19.6 GeV. It is important to note however that this alone does not guarantee that experimental data can be precisely fitted, as not necessarily all data points can be fitted correctly simultaneously.
\FloatBarrier
\begin{figure*}
\includegraphics{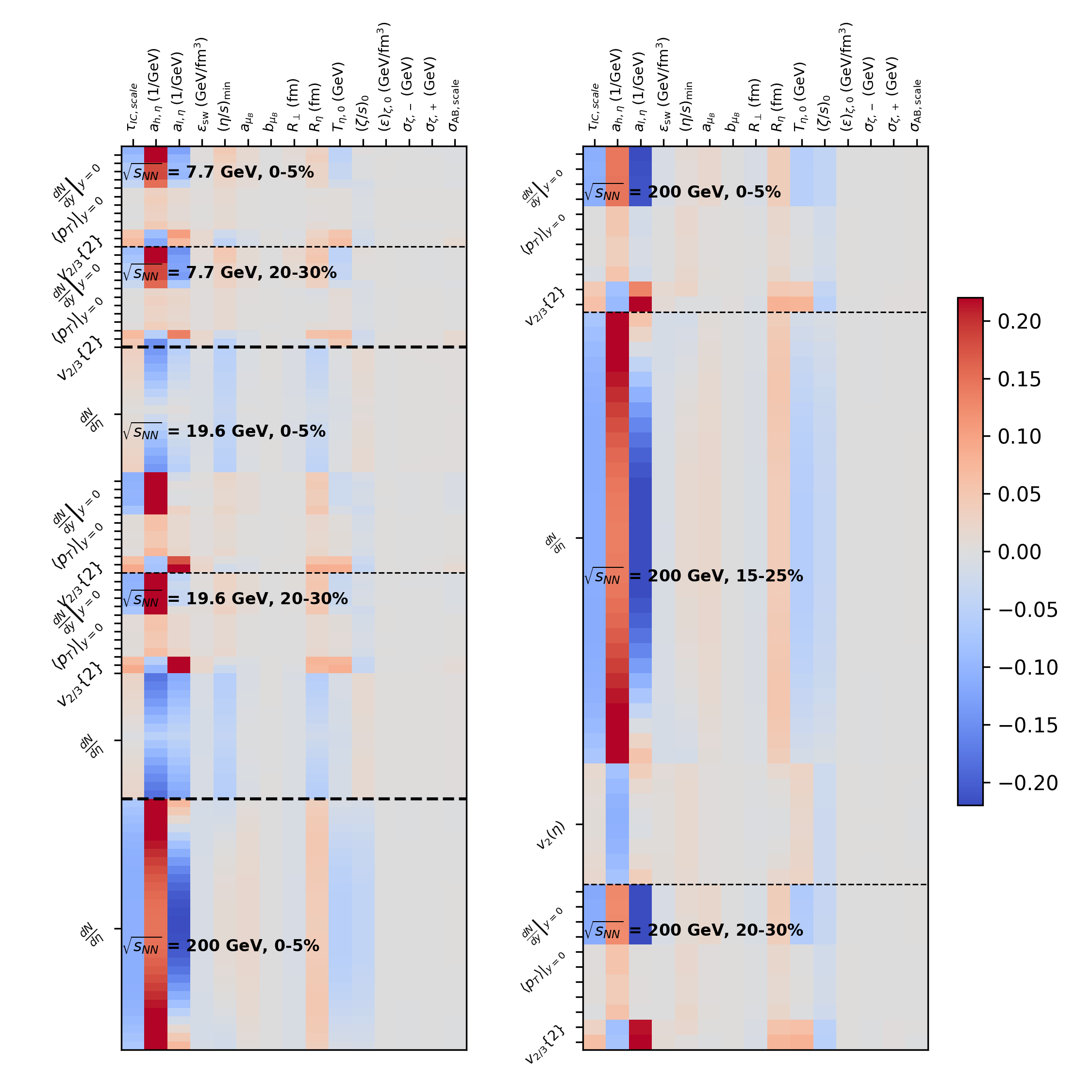}
\caption{Normalized partial derivatives of observables with respect to parameters of the simulation (response matrix) averaged over multiple prior points. Positive values (red) mark an increase in the observable with the value of the parameter, negative values (blue) mark a decrease. }
\label{fig:sensitivity}
\end{figure*}
\subsection{Closure test}
As a next step, we perform a closure test. Closure tests are crucial in validating that model parameters can be successfully constrained. They can, for example, detect if the number of points in prior space is too low, if data is insensitive to a parameter or if there are degeneracies in a model, allowing disjunct regions of the parameter space to result in the same predictions. Conducting a closure test follows a similiar idea than the emulator validation before. Again, one of the training points is separated out. However, this time the predictions of this training point are considered as pseudo-experimental data. Now, the whole toolchain, starting from the Gaussian process emulation and continuing to the construction  of the posterior using MCMC, is performed under the assumption of this pseudo-experimental data. In an ideal case, the original input parameters are reproduced, and the peaks of the posterior are close to these values. If the data is sufficient to constrain a parameter well, one expects strongly pronounced peaks.
\begin{figure}
    \centering
    \includegraphics{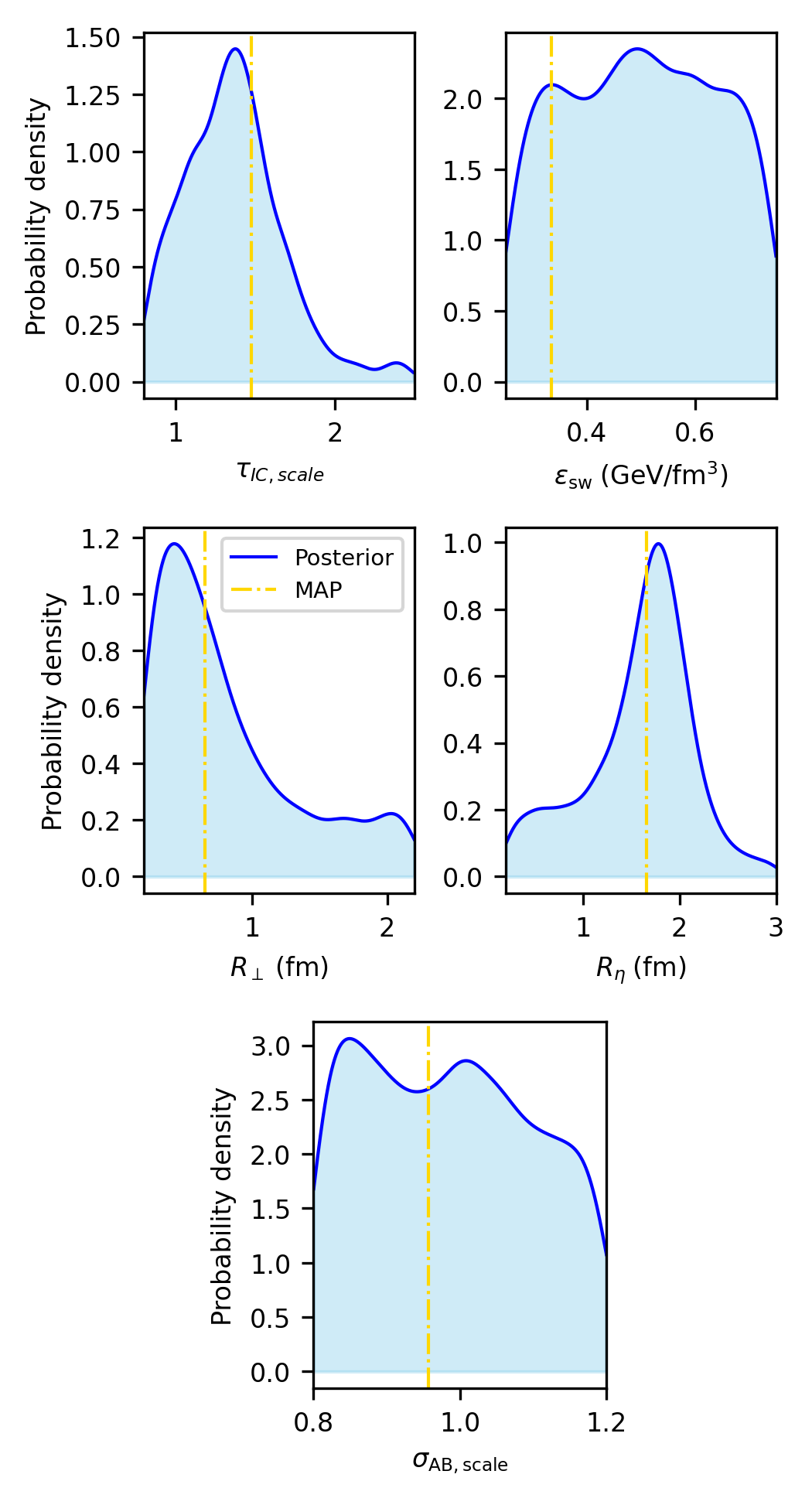}
    \caption{Posterior of the technical parameters. The golden vertical line represents the maximum-a-posteriori value.}
    \label{fig:full_tech}
\end{figure}
\begin{figure}
    \centering
    \includegraphics{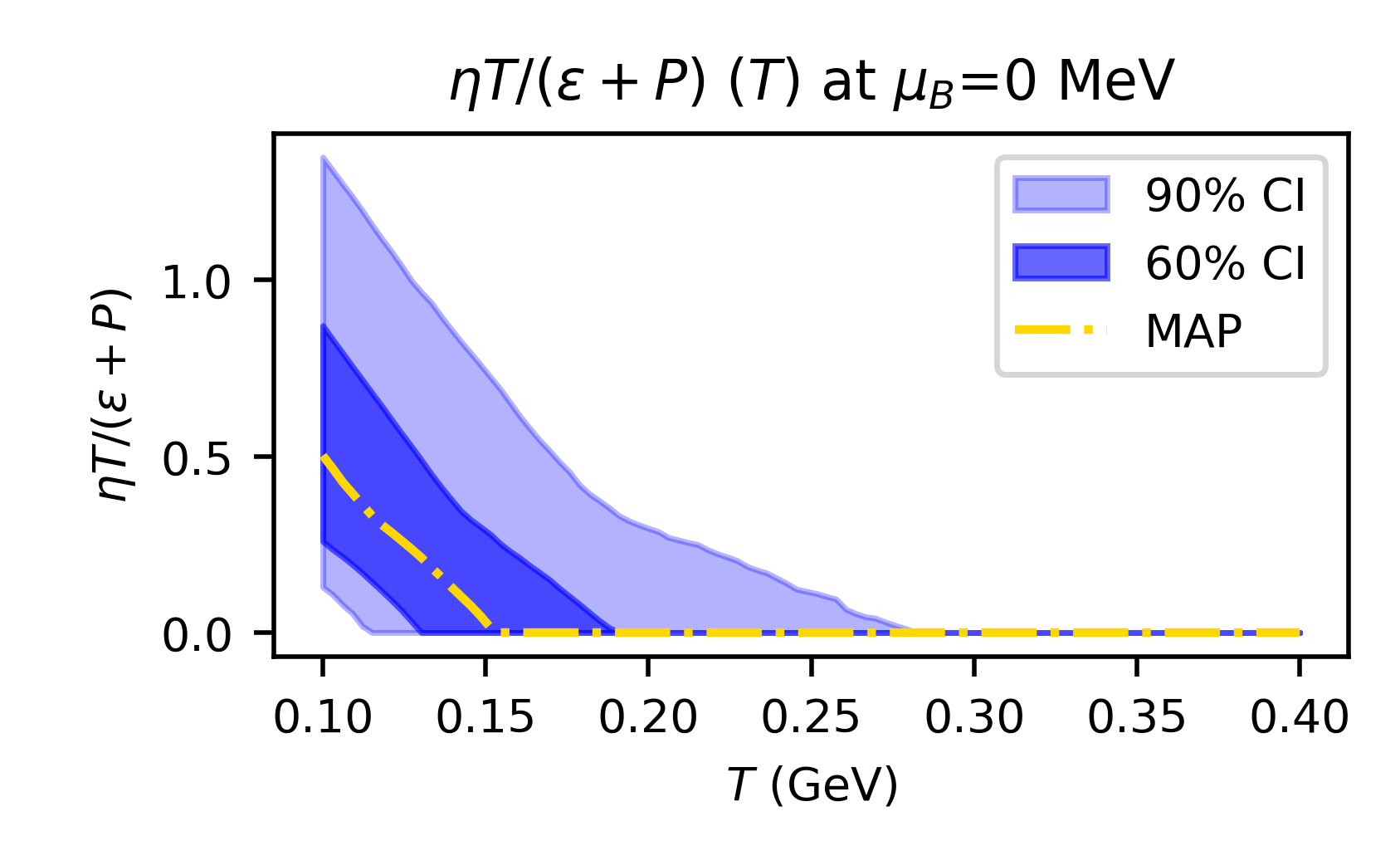}
    \includegraphics{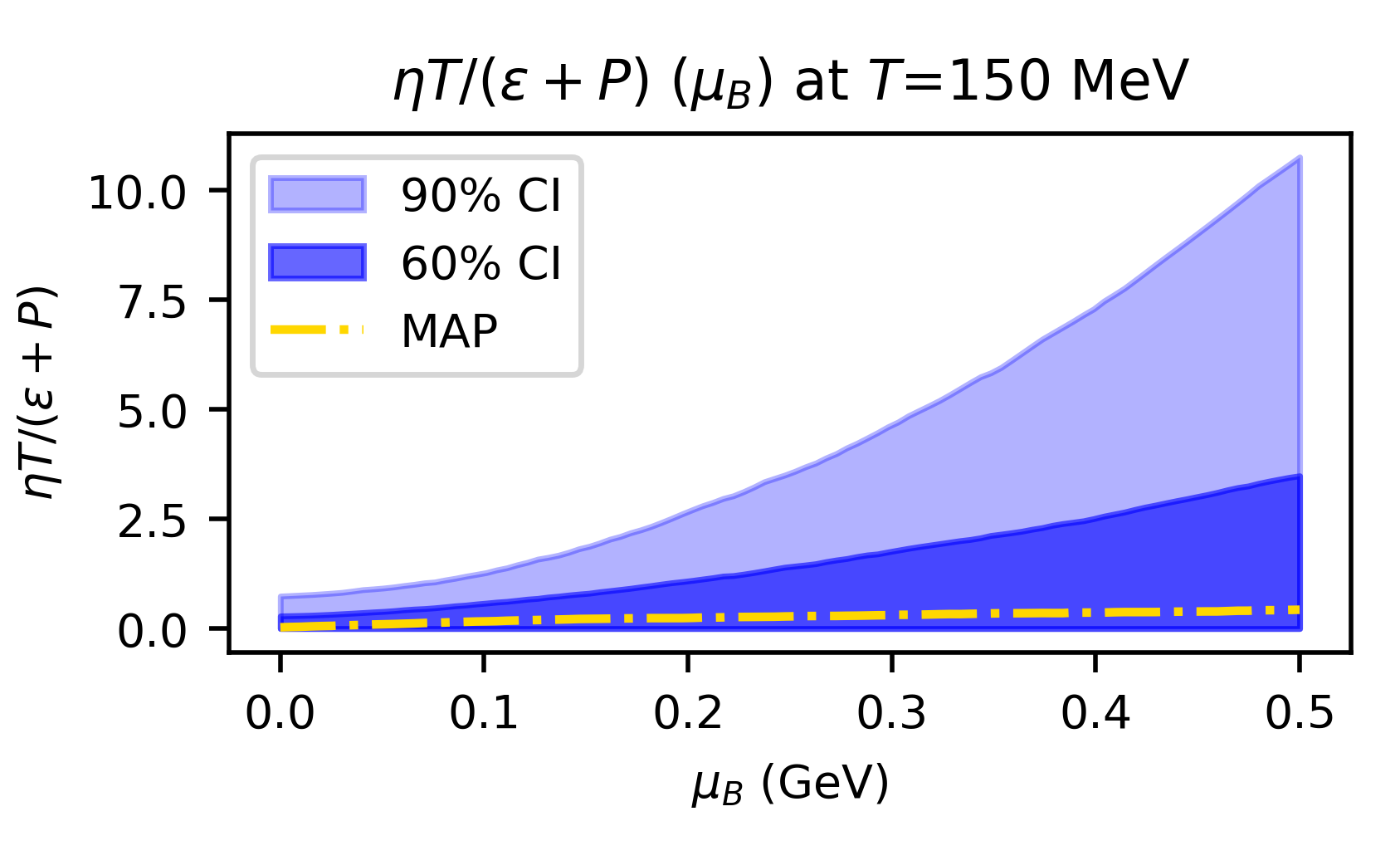}
    \includegraphics{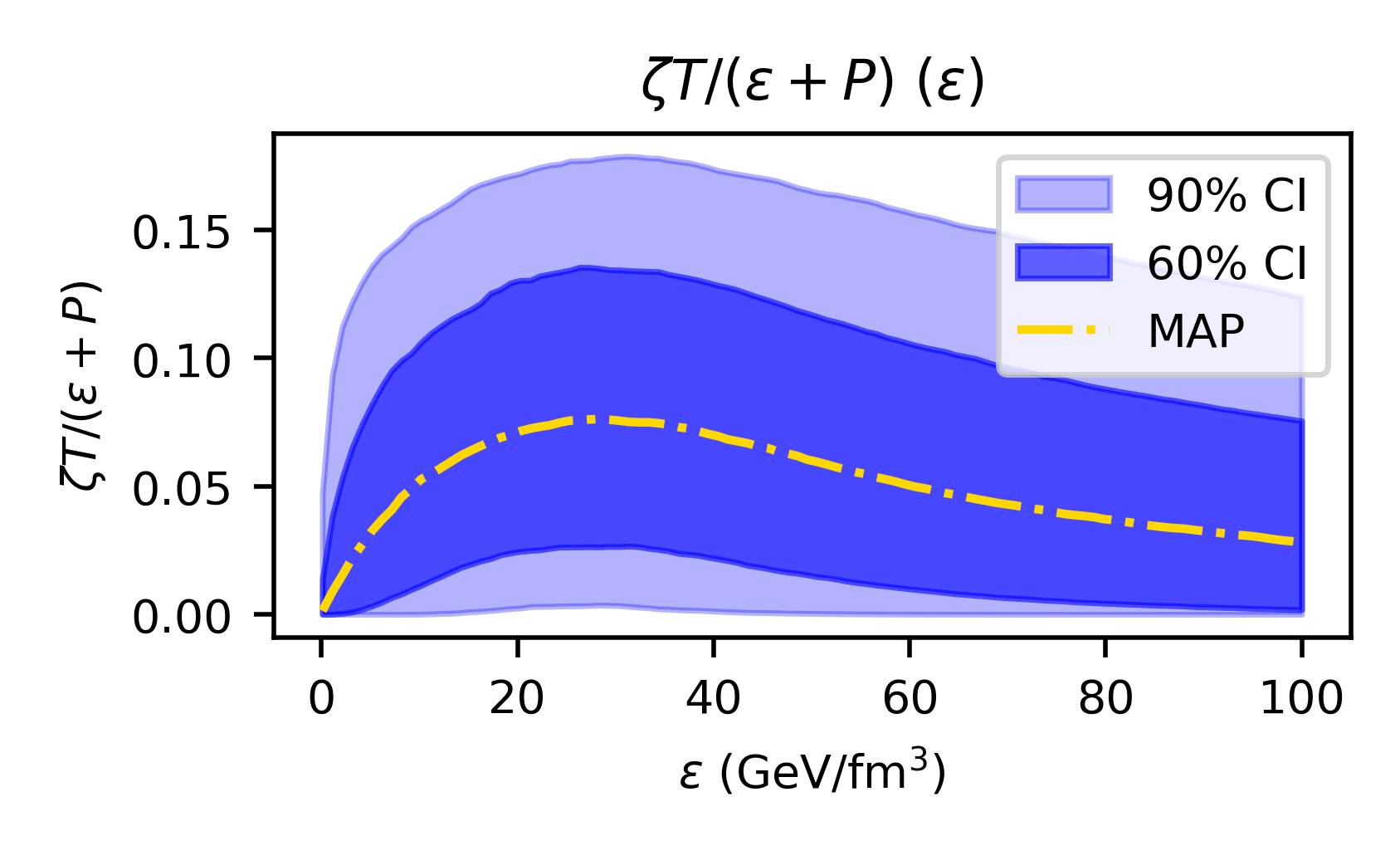}
    \caption{Posterior of the viscosities. From top to bottom: the shear viscosity as a  function of temperature for vanishing baryochemical potential, the shear viscosity as a function of baryochemical potential at fixed temperature, and the bulk viscosity as a function of the energy density. The golden line represents the maximum-a-posteriori. The bands represent the 60\% and 90\% confidence interval of the posterior.}
    \label{fig:full_visc}
\end{figure}
\Cref{fig:closure_tech} shows the posterior distributions of the technical parameters. Especially the scaling factor for the fluidization time and the smearing parameters are well constrained. The distribution is wider for the particlization energy density and especially for the late stage rescattering cross section scaling. Indeed, although for the latter, the MAP estimate agrees quite well with the original parameter, the distribution is very wide, showing only weak constraints.

Continuing to the viscosities in \cref{fig:closure_visc}, we see good constraints for the shear viscosity. Both as a function of temperature and as a function of baryochemical potential, the original parameterization lies comfortably in the 60\% confidence interval. This means that the Bayesian inference could extract the correct functional dependence. For the bulk viscosity, this holds in general, too. However, the quality of the posterior is here decreased and the original parameterization lies at the boundaries between the more strict 60\% confidence interval and the 90\% confidence interval. This is however still in the statistically acceptable range.

With this successful validation of our approach, we can now move on to the results on applying this setup to experimental data.

\begin{figure*}
    \centering
    \includegraphics[width=\linewidth]{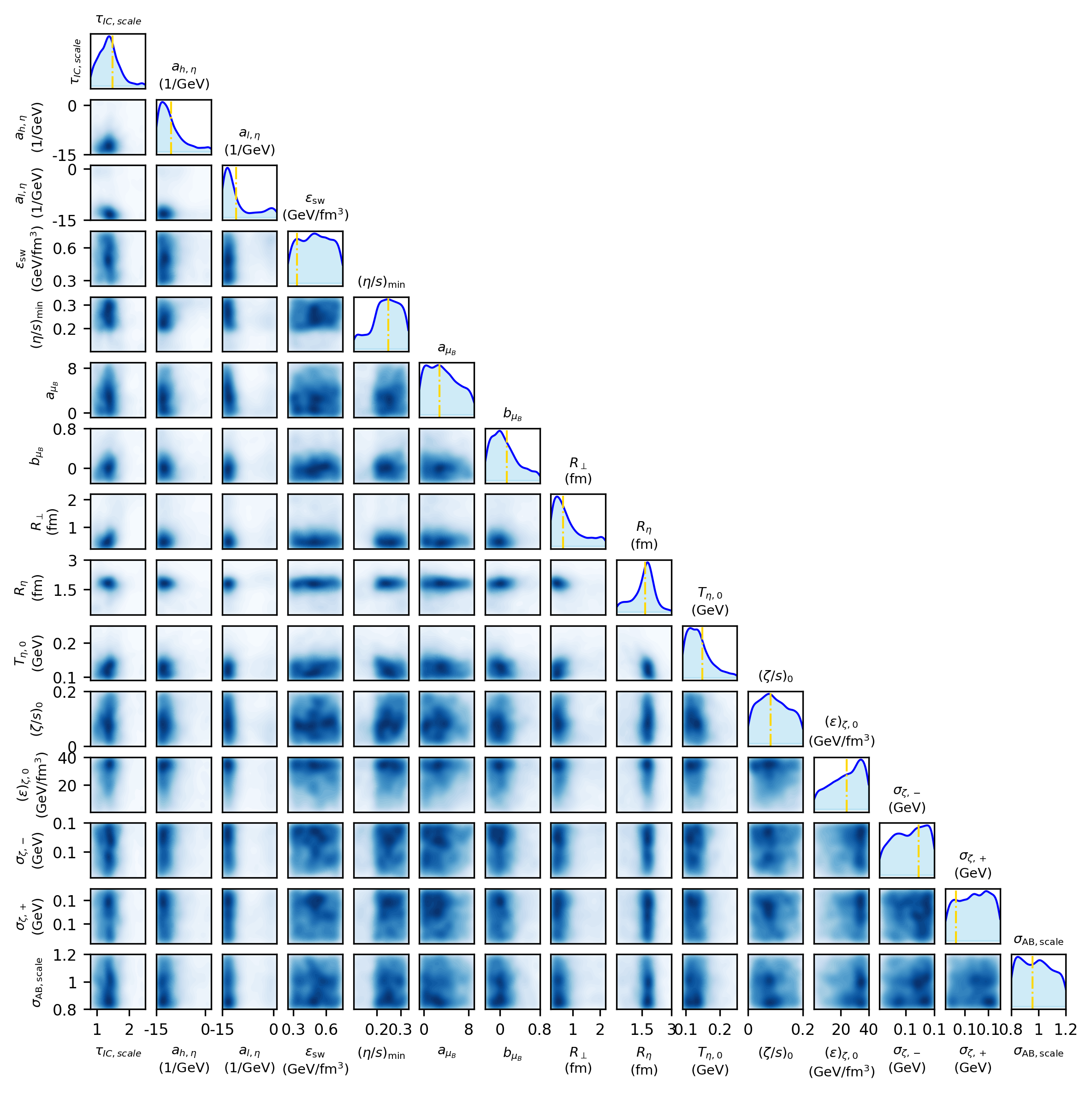}
    \caption{Full posterior for all parameters. The diagonal shows the distribution for each parameter, whereas the off-diagonal shows the probability distribution for a combination of two parameters.}
    \label{fig:triangle}
\end{figure*}
\section{Results}\label{sec:results}
We want to present now results obtained with our model.

\subsection{Sensitivity Analysis}
In a first step, we want to gain an understanding of how the different parameters affect observables. An easy accessible tool to do so is the response matrix, which looks at partial derivatives of observables with respect to the model parameters using the centered finite difference method. There are two important points to note about this: On the one hand, this is independent of the posterior, as this can be performed on the emulated prior directly. However, we perform this analysis in the center of the posterior. On the other hand, the response matrix is a local, linearized statement, and can change throughout the prior space, as there is potentially interaction between the parameters. Such relationships are contained in more evolved descriptions like Sobol indices.
\begin{figure*}
    \centering
    \includegraphics[width=\linewidth]{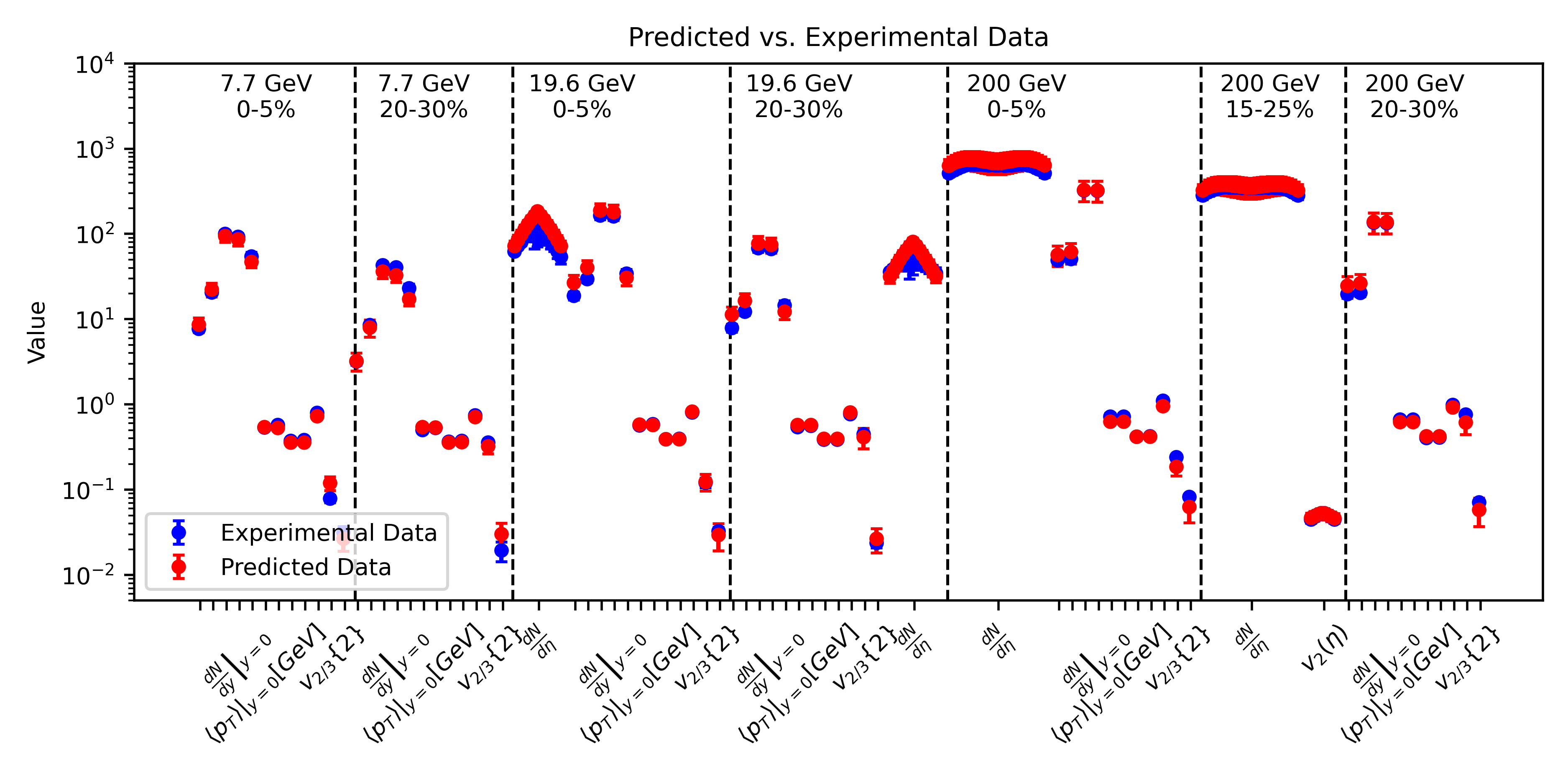}
    \includegraphics[width=\linewidth]{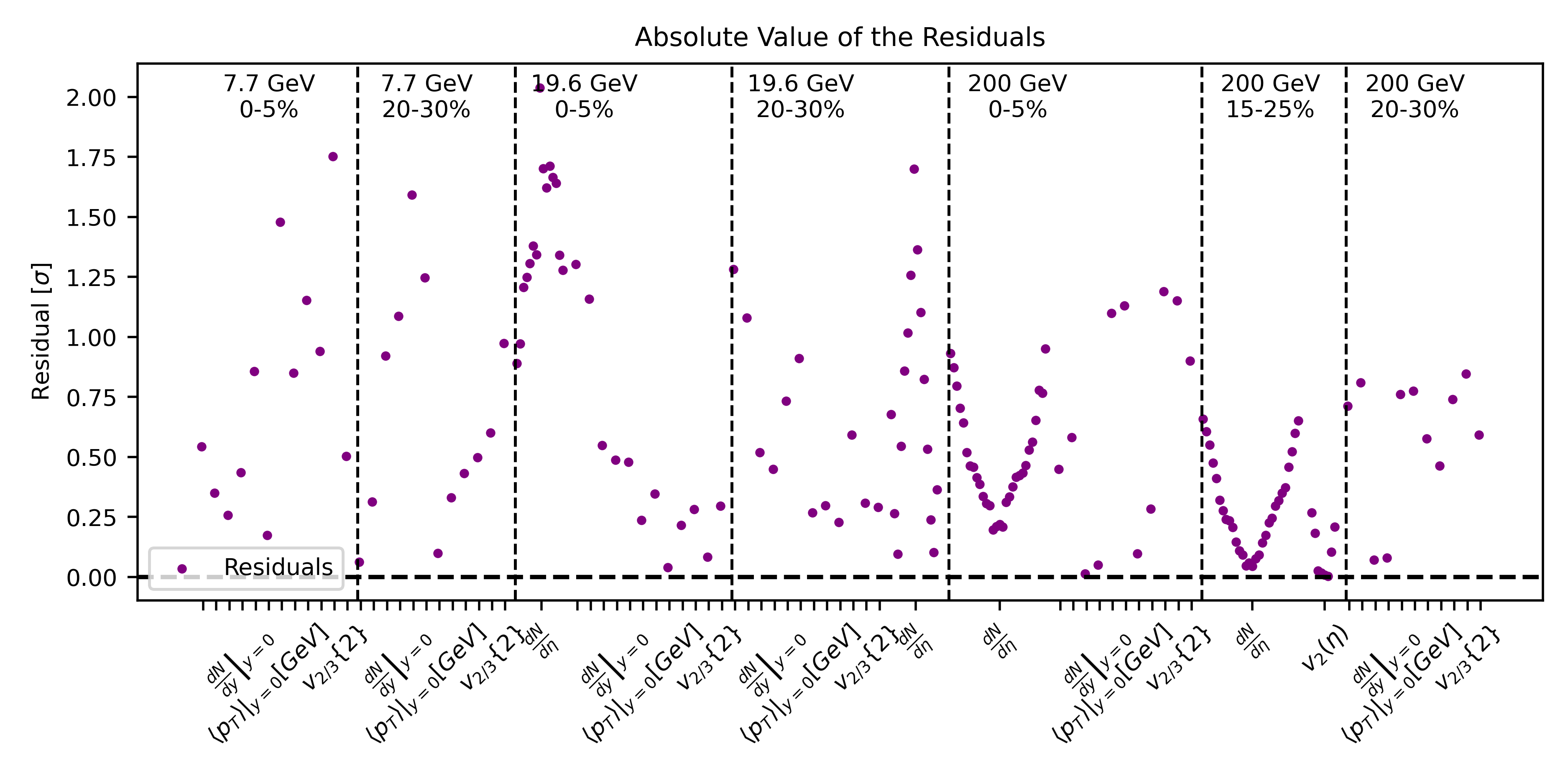}
    \caption{Comparison of predictions of observables from the posterior with experimental values (top). Residuals between predictions and experimental data in multiples of standard deviations (bottom).}
    \label{fig:datacomp}
\end{figure*}
To gain a first overview, we take a look at a normalized, averaged response matrix which aggregates the contributions of many prior points. The results are displayed in \cref{fig:sensitivity}. Positive (red) areas mean that a positive correlation between observable and parameter, negative (blue) means a negative correlation, i.e. a decrease in the parameter means an increase in the observable. We want to point out some striking observations.

The fluidization time scale $\tau_{\mathrm{IC, scale}}$ shows a high sensitivity in many observables. Most notably the flows, but also multiplicities. Whereas at midrapidity, one can observe consistently a negative correlation, at higher rapidities at intermediate energies, the fluidization time is positively correlated with multiplicities. This shows the complex effect of performing initial scatterings for a longer time. A possible explanation for this are differences in the treatment of particles in SMASH at higher and lower collision energies - at high energies, the particles remain generally unformed after the first collision, and cannot interact again, whereas secondary collisions are more likely at lower collision energies. We also note the positive correlation between flows and fluidization time scale; this is contradictory to earlier observed behaviour \cite{Karpenko:2015xea}. It suggests that early transport is more efficient in generating momentum anisotropy than the hydrodynamic stage, which could be an effect of the presence of substantial values of the transport coefficients.
The longitudinal smearing increases multiplicities at high energies while decreasing it in the high rapidity regions at intermediate energies. This gives a more nuanced picture on their effect for the multiplicities with respect to \cite{Karpenko:2015xea}.
Another important parameter is the kink temperature of the shear viscosity, $T_{\eta,0}$. It shows in general a similiar behaviour to the fluidization time scaling, although it decreases multiplicities in the full rapidity spectrum at intermediate energies. The steepnesses of the linear temperature dependence of the shear viscosity show strong impacts on a wide range of observables, with a substantial presence of rapidity dependence. Note that the correlations are of opposite sign due to the additional minus sign for low temperatures in \cref{eq:shear}. In general, the expected behavior is realised that increasing the shear viscosity decreases anisotropic flows. There is only weak dependency on parameters related to baryochemical potential. For bulk viscosity, only the parameter of its magnitude is relevant. All other viscous parameters do not affect the observables strongly. The same can be said about the rescattering cross section scaling and the density dependence of the shear viscosity, which show relatively small signals. Therefore, we can expect weaker constraints on these parameters.

\begin{figure}
    \centering
    \includegraphics{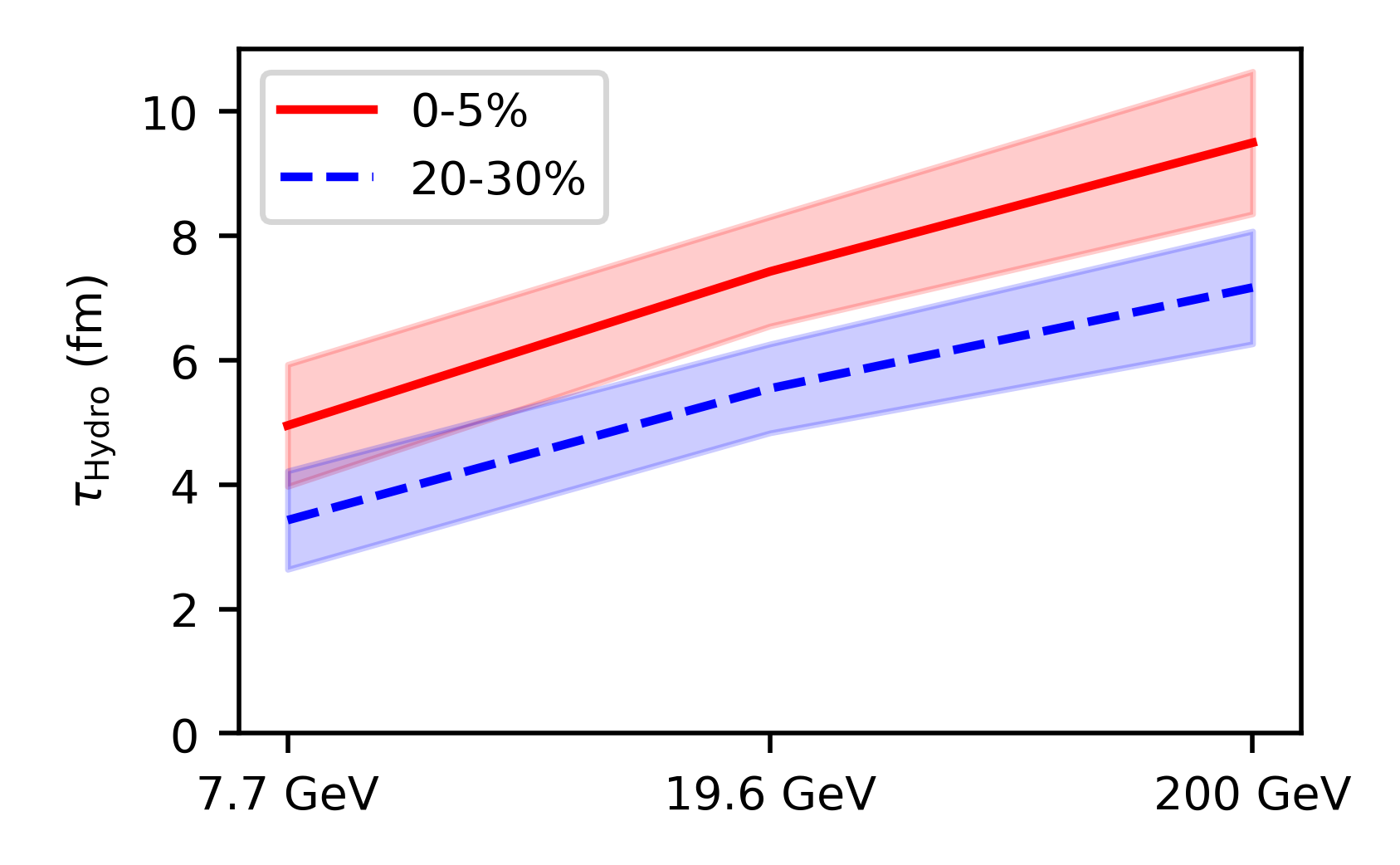}
    \caption{MAP of the duration of the hydrodynamic evolution in the posterior.}
    \label{fig:hydrotime}
\end{figure}

\begin{figure*}
    \centering
    \includegraphics{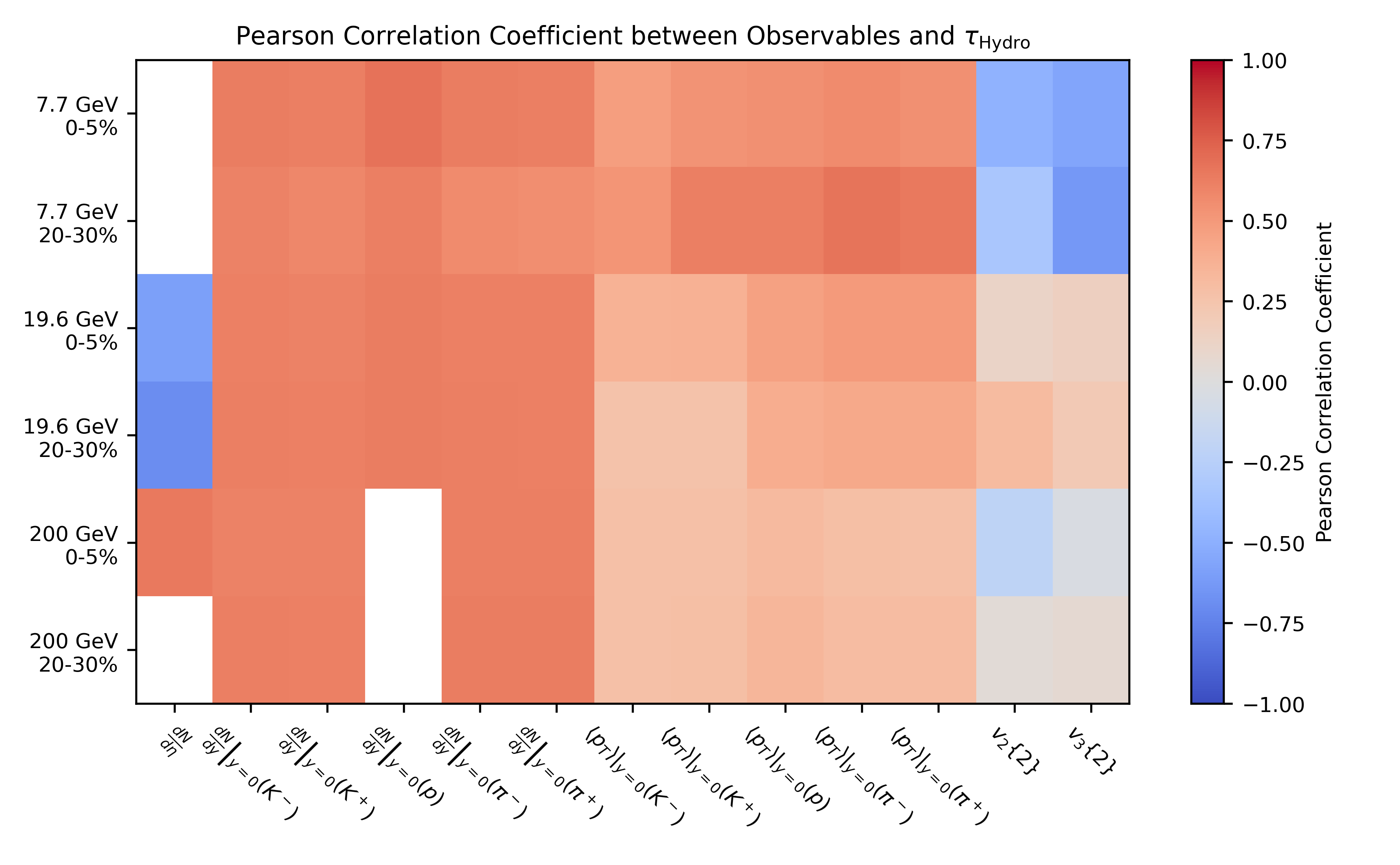}
    \caption{Pearson correlation coefficient between the hydrodynamic evolution duration and observables used to train the fit. White squares signify observables where were not used at the respective energies/centralities. For spectra, the average over the data points is displayed.}
    \label{fig:hydrotime_corr}
\end{figure*}

\subsection{Posterior Predictions}
In a first step, we take a look  at the posterior distribution of technical parameters again, which can be found in \cref{fig:full_tech}.  We observe a preference for slightly increased fluidization time, which contrasts with the results of
~\cite{Auvinen:2017fjw} preferring a minimal fluidization time. Addtionally, we find a lower particlization energy density of around 0.33 $\frac{\mathrm{GeV}}{\mathrm{rm}^3}$ to be optimal, albeit at substantial uncertainty. The preferred smearing parameters differ substantially between both studies. The current model prefers minimal smearing in the transverse plane and substantial smearing in the longitudinal direction.  Regarding the scaling of cross sections in the late stage rescattering, the uncertainty is high and therefore the result is statistically consistent with 1.

\begin{figure}
    \centering
    \includegraphics{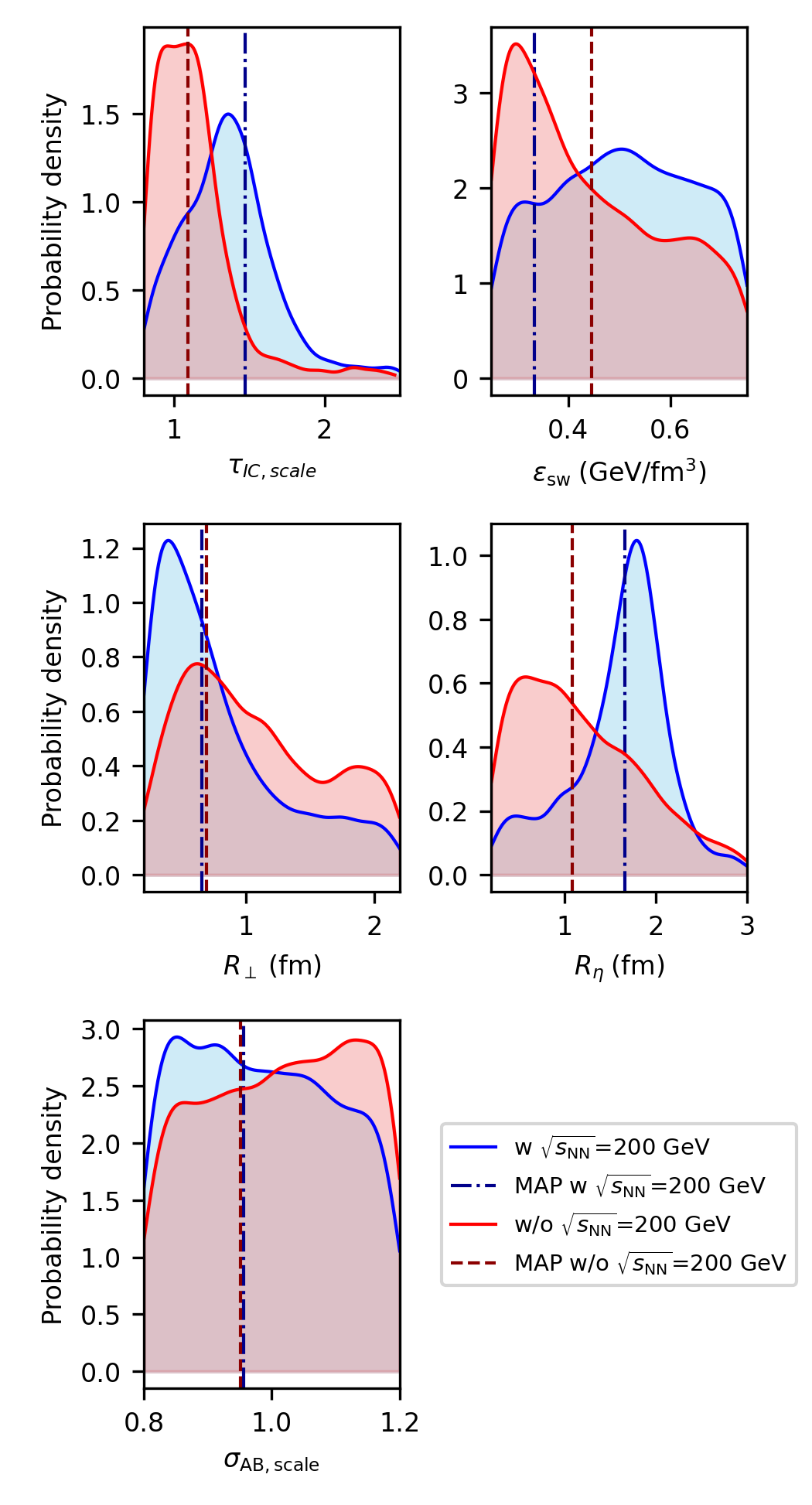}
    \caption{Posterior of the technical parameters when including and excluding data at $\sqrt{s_{\mathrm{NN}}}$= 200 GeV.}
    \label{fig:energy_tech}
\end{figure}

\begin{figure}
    \centering
    \includegraphics{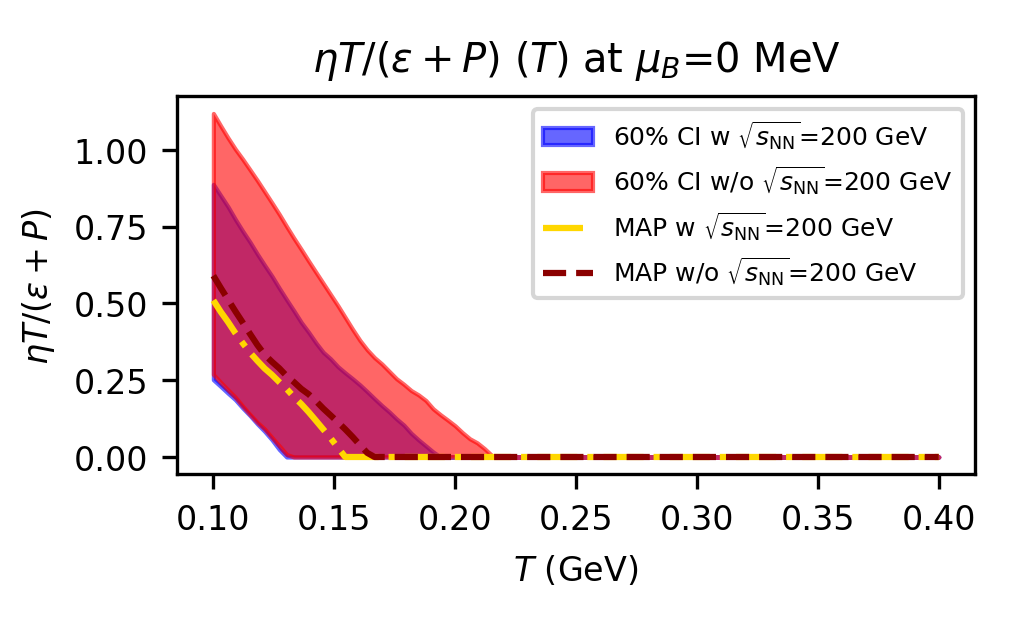}
    \includegraphics{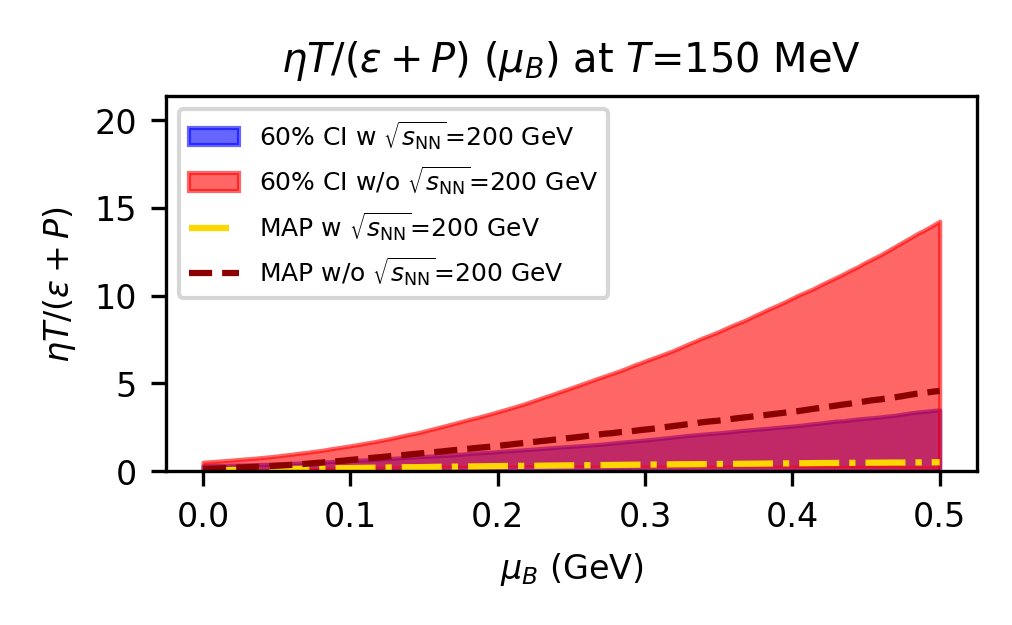}
    \includegraphics{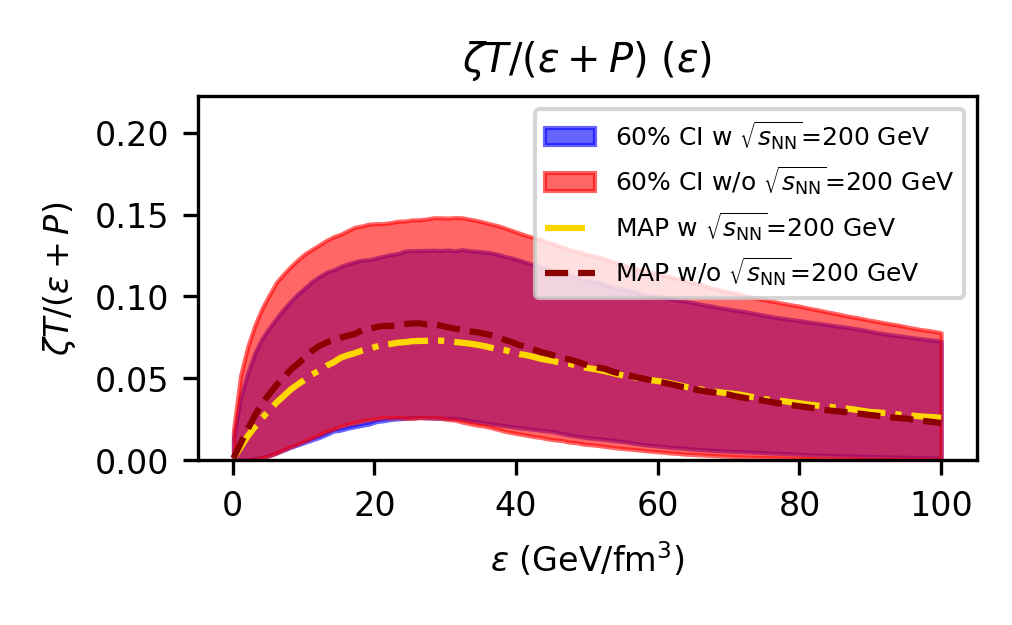}
    \caption{Posterior of the viscosities when including and excluding data at $\sqrt{s_{\mathrm{NN}}}$= 200 GeV. From top to bottom: the shear viscosity as a  function of temperature for vanishing baryochemical potential, the shear viscosity as a function of baryochemical potential at fixed temperature, and the bulk viscosity as a function of the energy density. The bands represent the 60\%  confidence interval of the posterior.}
    \label{fig:energy_visc}
\end{figure}
Looking at viscosities in \cref{fig:full_visc}, the observations drawn earlier from the seen for the sensitivity analysis become apparent again: the constraints on both the bulk viscosity and the baryochemical potential dependence are too weak to efficiently constrain these values and a wide range of values is consistent with the data.

\begin{figure}
    \centering
    \includegraphics{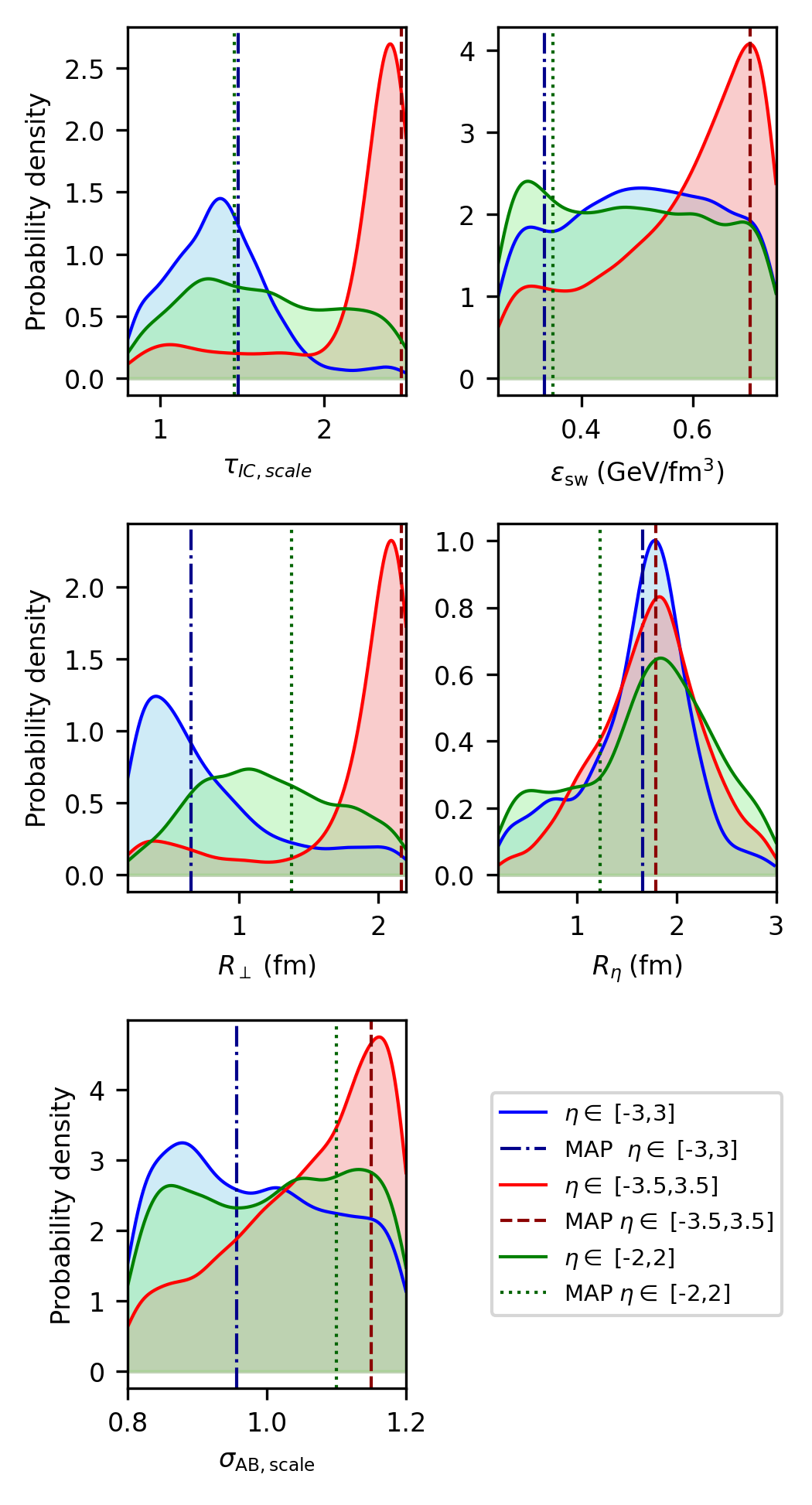}
    \caption{Posterior of the technical parameters depending on the rapidity cut.}
    \label{fig:rapidity_tech}
\end{figure}

\begin{figure}
    \centering
    \includegraphics{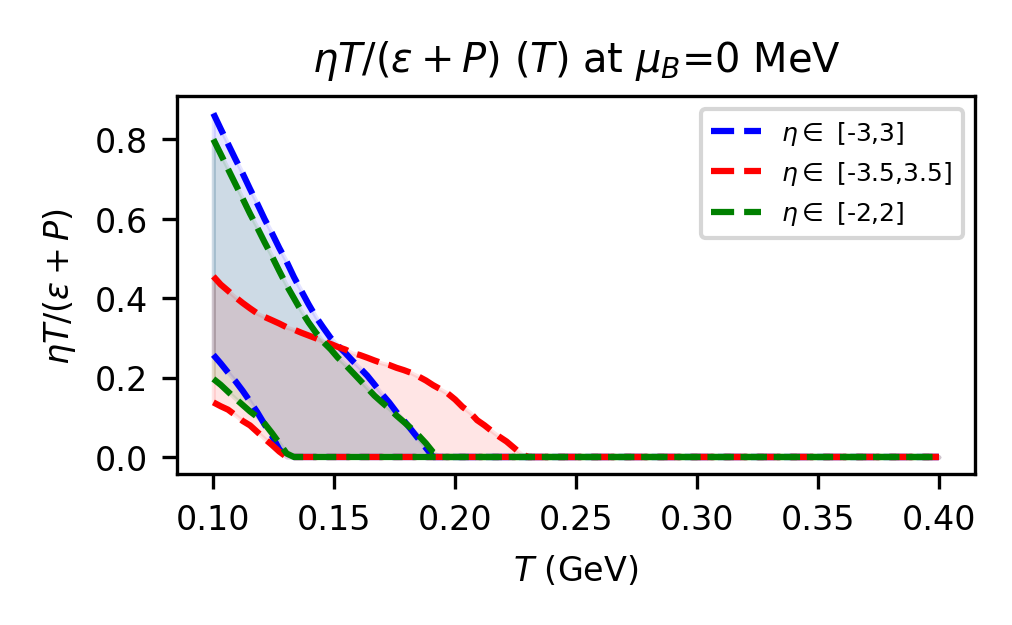}
    \includegraphics{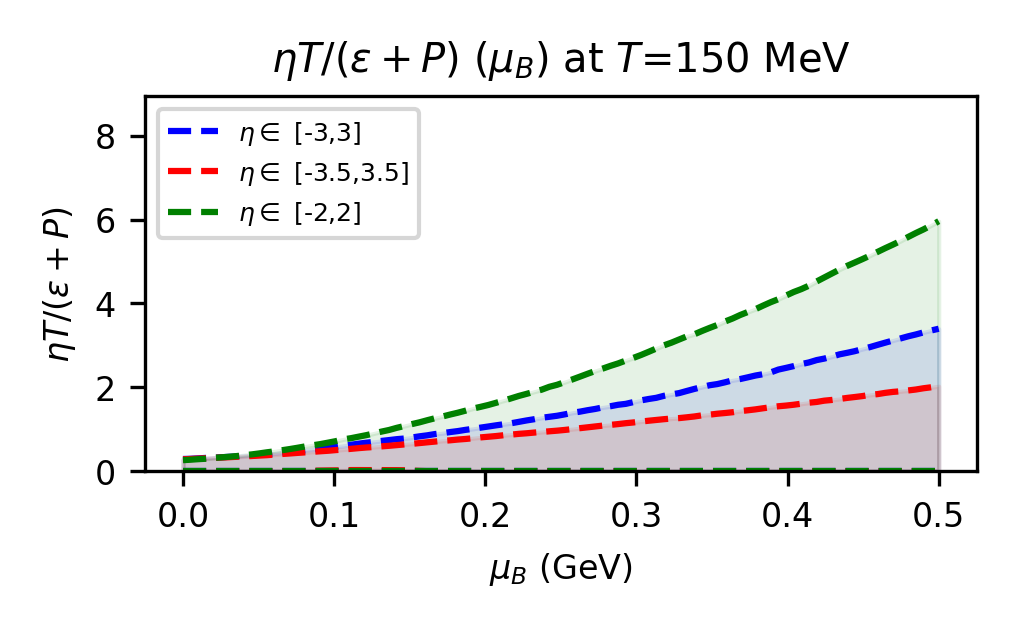}
    \includegraphics{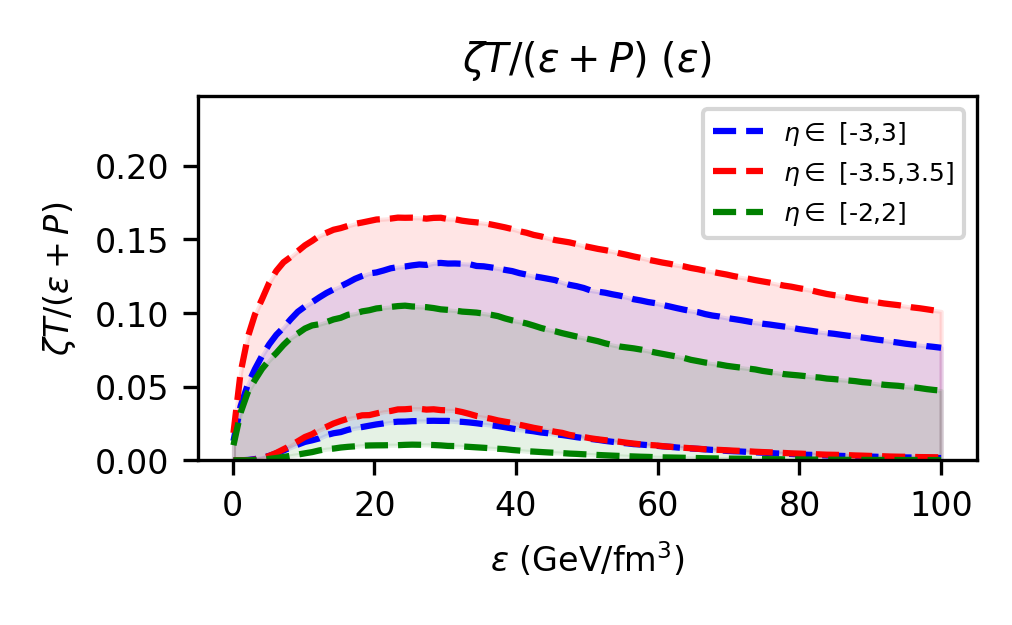}
    \caption{Posterior of the viscosities depending on the rapidity cut. From top to bottom: the shear viscosity as a  function of temperature for vanishing baryochemical potential, the shear viscosity as a function of baryochemical potential at fixed temperature, and the bulk viscosity as a function of the energy density. The bands represent the 60\%  confidence interval of the posterior.}
    \label{fig:rapidity_visc}
\end{figure}

The situation is different for the temperature dependence, where a rapidly decreasing shear viscosity is preferred, which stays at zero from around 150 - 250 MeV on. The origin of this becomes more clear looking at the full posterior of all parameters in \cref{fig:triangle}. We can see that, although there is a kink with $(\eta/s)_{\mathrm{min}} \ne 0$, the steepness on both sides is nearly identical, leading to the cutoff to negative values setting the shear viscosity to 0 for high temperatures. Interestingly, this is a degenerated form - after all, a kink with $(\eta/s)_{\mathrm{min}}$ = 0 and  any negative steepness for higher temperatures could reach such a configuration. It is currently not clear why such a setup was preferred. Nevertheless, it seems that our model strongly prefers vanishing shear viscosities in the QGP phase, but substantial viscosities near particlization. All maximum-a-posteriori values can be also found in \cref{tab:map_parameter}.

There are further insights to be gained from \cref{fig:triangle}. We see that $b_{\mu_B}$ is strongly peaked at 0. Therefore, much of the uncertainty for baryochemical potential dependence comes from $a_{\mu_B}$, which is insufficiently constrained.
 Furthermore, we see only very limited interactions between the parameters, which can be most easily identified by diagonal structures in the triangle plot. A trace of this can be seen for $\tau_{\mathrm{IC,scale}}$ and $R_\perp$.

We want to point out that our tune reproduces experimental data at a very high quality. In \cref{fig:datacomp}, we show a comparison between the experimental data and the predicted values from the model, which include the residual uncertainty in the parameters. For all data points, the difference is around 2 standard deviations or lower. It is interesting to note that with regards to the longitudinal properties, at high energies, our model describes midrapidity multiplicities better than data at forward or backward rapidities, whereas at intermediate energies, this is inversed. This points to a general limitation of this model to fully capture longitudinal dynamics as a function of collision energies. The comparison of the MAP observable predictions with experimental measurements is also provided in a linear scale and split by observables in the appendix in \cref{fig:datacomp_linear_bulk}, \cref{fig:datacomp_linear_flow} and \cref{fig:datacomp_linear_eta}.

\subsection{Hydrodynamic evolution time}
Both $\tau_{\mathrm{IC, scale}}$ and $\epsilon_{\text{switch}}$ encode the duration of the hydrodynamic evolution, as they determine the starting and ending point, respectively. This motivates the study of $\tau_{\mathrm{Hydro}}$, the time which is spent in the hydrodynamic evolution. This time is also very crucial for the fit as it is this time frame of the evolution where viscosities can be tuned for modifying the viscosity.\Cref{fig:hydrotime} shows the maximum a posteriori for the duration of the hydrodynamic evolution for different systems. As expected, it increases both with collision energy and centrality. The values roughly agree with the ones reported for the default setup in \cite{Schafer:2021csj}.

This is no coincidence as $\tau_{\mathrm{Hydro}}$ has a strong effect on observables. \Cref{fig:hydrotime_corr} shows the correlation between observables and the hydrodynamic evolution time. Especially for yields and for flows at low energies, the correlation is significant. The reason for the flows being affected stronger at low energies can be attributed to the medium of lower temperature and higher baryochemical potential, which yields higher viscosities.

Nevertheless, $\tau_{\mathrm{Hydro}}$ can  not be seen as a parameter of the system, replacing $\tau_{\mathrm{IC, scale}}$ and $\epsilon_{\text{switch}}$. Increasing both of these parameters to yield a similiar average $\tau_{\mathrm{Hydro}}$ would not amount to the same observables, as the evolution of the system in the early and late stages is fundamentally different. In other words, the hydrodynamic evolution can not be shifted throughout the evolution of the heavy ion collision. This can be seen also from the data directly: \cref{fig:triangle} shows no diagonal structures between $\tau_{\mathrm{IC, scale}}$ and $\epsilon_{\text{switch}}$. Therefore, there are no corresponding pairs between these parameters leading to similiar results.

\subsection{Impact of data selection}

We want to take a deeper look  on the role of longitudinal data and the collision energies investigated. In a first step, we exclude data at $\sqrt{s_{\mathrm{NN}}}$= 200 GeV. \Cref{fig:energy_tech} shows the effect on the technical parameters. While there is only a small change for the cross section scaling and the transverse smearing, we observe substantial changes for other technical parameters. The initial stage scaling is now centered around 1. Most notably, however, is the preference for increased lifetime of the hydrodynamic stage and a reduction in longitudinal smearing. This strong change of parameters was most likely pushed by the residuals in multiplicity, which in the full fit were still substantial at intermediate energies, due to balancing out with the residuals at high rapidity data from high energies. Investigating further the viscosities, \cref{fig:energy_visc} shows that data at high energies gives further constraints. Although the shear viscosity as a function of baryochemical potential is stronger constrained, this is a combination of two contributions: on the one hand, the additional rapidity-dependent data at 200 GeV contributes to the constraints. On the other hand, the additional data also constraints the temperature dependence, leading to a more constraint viscosity at $T= 150$ MeV at vanishing baryochemical potential. As this is scaled in proportion to the baryochemical potential, a more peaked temperature dependence also gives a less wide distribution as a function of $\mu_B$.

The crucial role played by the longitudinal data is further investigated in \cref{fig:rapidity_tech} and \cref{fig:rapidity_visc}. Here, we compare the full dataset, but increase and reduce the rapidity cut. Regarding the technical parameters, cutting more strictly does not statistically significantly change most parameters, except increasing slightly the transversal smearing. For the relaxed cut, on the other hand, we see multiple of the technical parameters running into the boundaries of the prior ranges. We can therefore conclude that with our setup, we can not appropriately describe data at such high rapidities.

Regarding viscosities, we see that including high rapidity data gives additional constraints on the baryochemical potential dependence. However, it increases uncertainties of the bulk viscosity.


\section{Conclusions and Outlook}\label{sec:Conclusion}

In this work, we have performed a Bayesian analysis of the (3+1)D SMASH-vHLLE-hybrid model across a range of collision energies from 7.7 to 200 GeV. By comparing various sets of observables to a broad parameter space, we have constrained both the technical parameters (initialization time, transverse/longitudinal smearing, and switching energy density) and the temperature- and baryochemical-potential-dependent viscosities of the QGP. We have shown that, with the data currently available, the model strongly favors a near-zero shear viscosity in the high-temperature phase while permitting moderate-to-large values in the vicinity of the pseudo-critical region. Moreover, a wide range of bulk viscosity values remain compatible with the data, and the preferred baryochemical potential dependence is essentially consistent with no dependence at all, emphasizing the limitation in present data precision.

These findings, especially regarding the nearly vanishing specific shear viscosity at higher temperatures, are at odds with several existing Bayesian analyses~\cite{Bernhard:2019bmu,JETSCAPE:2020shq,Parkkila:2021yha}, which generally support a non-zero minimal shear viscosity in the QGP phase. We attribute our discrepancy partly to the reduced freedom in the initial state provided by the hadronic transport model, which lowers the degrees of freedom for tuning geometric fluctuations and thus pushes certain parameters, such as the hydrodynamic onset time, into especially sensitive roles. While our analysis accommodates key beam energies and rapidity-dependent observables, even higher-statistics data would be needed for more definitive conclusions. A natural next step is to simplify the modelling assumptions by removing the presently unconstrained baryochemical potential dependence in order to reduce the dimensionality of the parameter space and isolate the most dominant effects. Similarly, it is a worthwhile study to investigate how much of the strong constraints on the high temperature shear viscosity can be attributed to the presence of bulk viscosity. We also envision direct comparisons to other initial-state scenarios, such as \textsc{TRENTo}, UrQMD, or McDIPPER, to examine the interplay between initial conditions and viscosity extractions. Ultimately, our investigation highlights the continued need for detailed experiments and improved multi-stage modeling to achieve a more robust picture of the QGP’s transport properties over a wide range of beam energies.

Raw data of figures can be downloaded from the supplemental material \cite{supp}.

\begin{acknowledgments}
This work was supported by the Deutsche Forschungsgemeinschaft (DFG, German Research
Foundation) – Project number 315477589 – TRR 211. N.G. acknowledges support by the Stiftung Polytechnische Gesellschaft Frankfurt am Main as well as the Studienstiftung des Deutschen Volkes. I.K. acknowledges support by the Czech Science Foundation under project No.~25-16877S. Computational resources have been provided by the GreenCube at GSI. N.G. wants to thank Hendrik Roch for the continued discussion and advise throughout this work.
\end{acknowledgments}


\appendix
\section{MAP of parameters}
The following table shows the maximum-a-posteriori of the parameters, equivalent to the golden lines in \cref{fig:triangle}, with 1-$\sigma$-confidence interval.
\begin{table}[h!]
    \centering
    \renewcommand{\arraystretch}{1.5}
    \begin{tabular}{c|c c}
        \hline \hline
        \textbf{Parameter} & \textbf{MAP} &\\ \hline
         $R_{\perp}$           & $0.6535^{+0.7536}_{-0.6582} $ & fm \\ 
        $R_{\eta}$            & $1.6577^{+0.3906}_{-0.6767}$ & fm \\ 
        $\tau_{\text{IC,scale}}$              & $1.4741^{+0.1703}_{-0.4266}$&  \\ 
        $\epsilon_{\text{switch}}$ & $0.3339^{+0.3298}_{-0.0233}$ & $\frac{\text{GeV}}{\text{fm}^3}$ \\ 
        $a_{l,\eta}$           & $-10.9149^{+6.5179}_{-3.2712}$ & $\frac{1}{\text{GeV}}$ \\ 
        $a_{h,\eta}$           & $-10.5116^{+3.3021}_{-3.3842} $& $\frac{1}{\text{GeV}}$\\ 
        $T_0$           & $0.1480^{+0.0223}_{-0.0435}$ & GeV  \\ 
        $(\eta/s)_{\text{min}}$ & $0.2214^{+0.0806}_{-0.0931}$ & \\ 
        $a_{\mu_B}$     & $2.8399^{+3.8013}_{-2.4225}$ & \\ 
        $b_{\mu_B}$     & $0.1379^{+0.7536}_{-0.3114} $& \\ 
         $\zeta_0$           & $0.0816^{+0.0785}_{-0.0483} $& \\ 
        $\epsilon_\zeta$    & $24.0359^{+11.7138}_{-13.3645} $& $\frac{\text{GeV}}{\text{fm}^3}$ \\ 
        $\sigma_{\zeta,-}$  & $0.0725^{+0.0154}_{-0.0474}$ & \\ 
        $\sigma_{\zeta,+}$  &$ 0.0378^{+0.0913}_{-0.0025} $& \\
        $\sigma_{\rm AB, \rm scale}$ & $0.9565^{+0.1639}_{-0.10484}$ & \\ \hline \hline
    \end{tabular}
    \caption{MAP of all parameters of the model.}
    \label{tab:map_parameter}
\end{table}

\section{MAP observable predictions}
In the following, we show the maximum a posterior observable predictions in comparison to experimental data. References for the experimental data can be found in \cref{table:expData}.

\begin{figure}
    \centering
    \includegraphics[width=\linewidth]{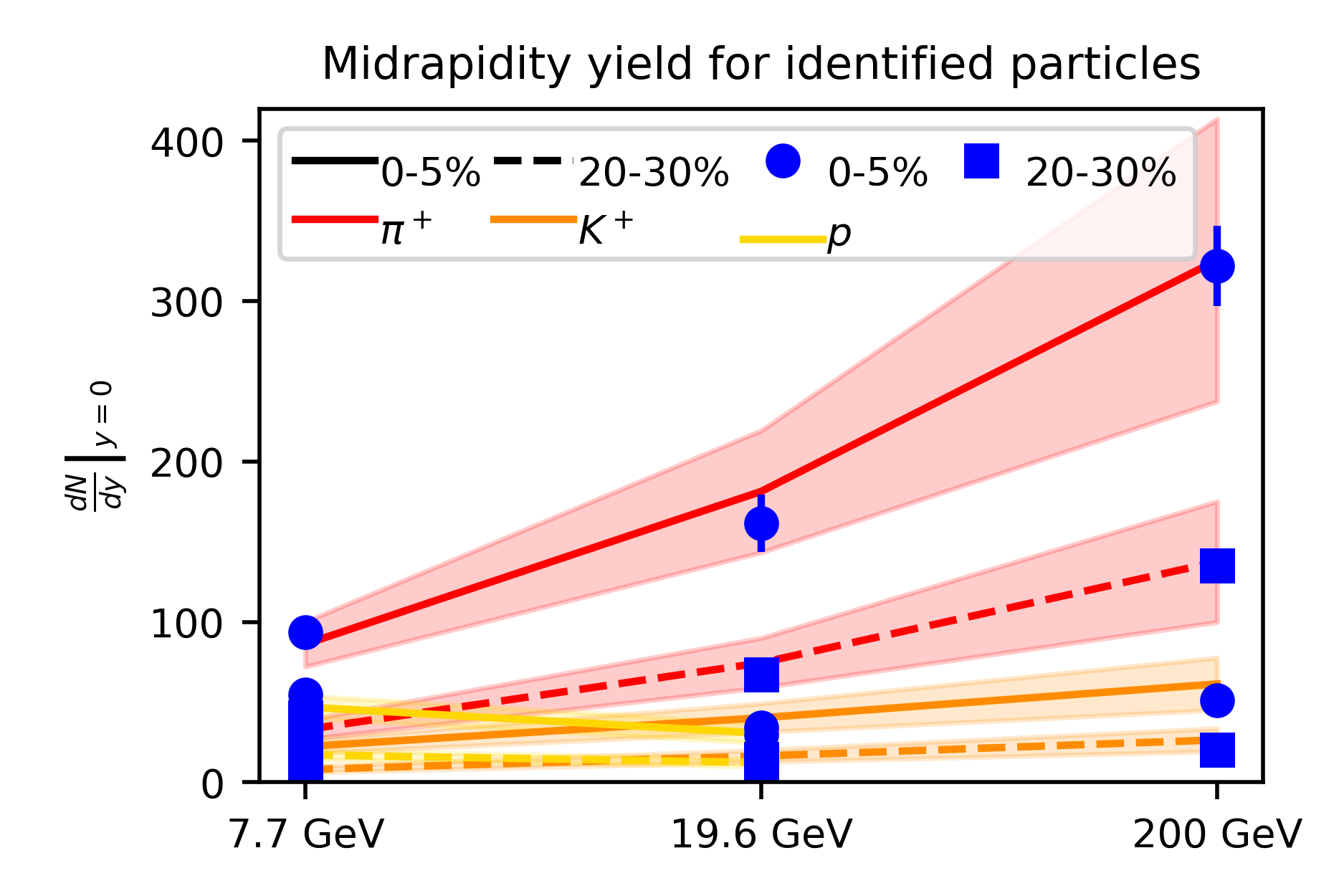}
    \includegraphics[width=\linewidth]{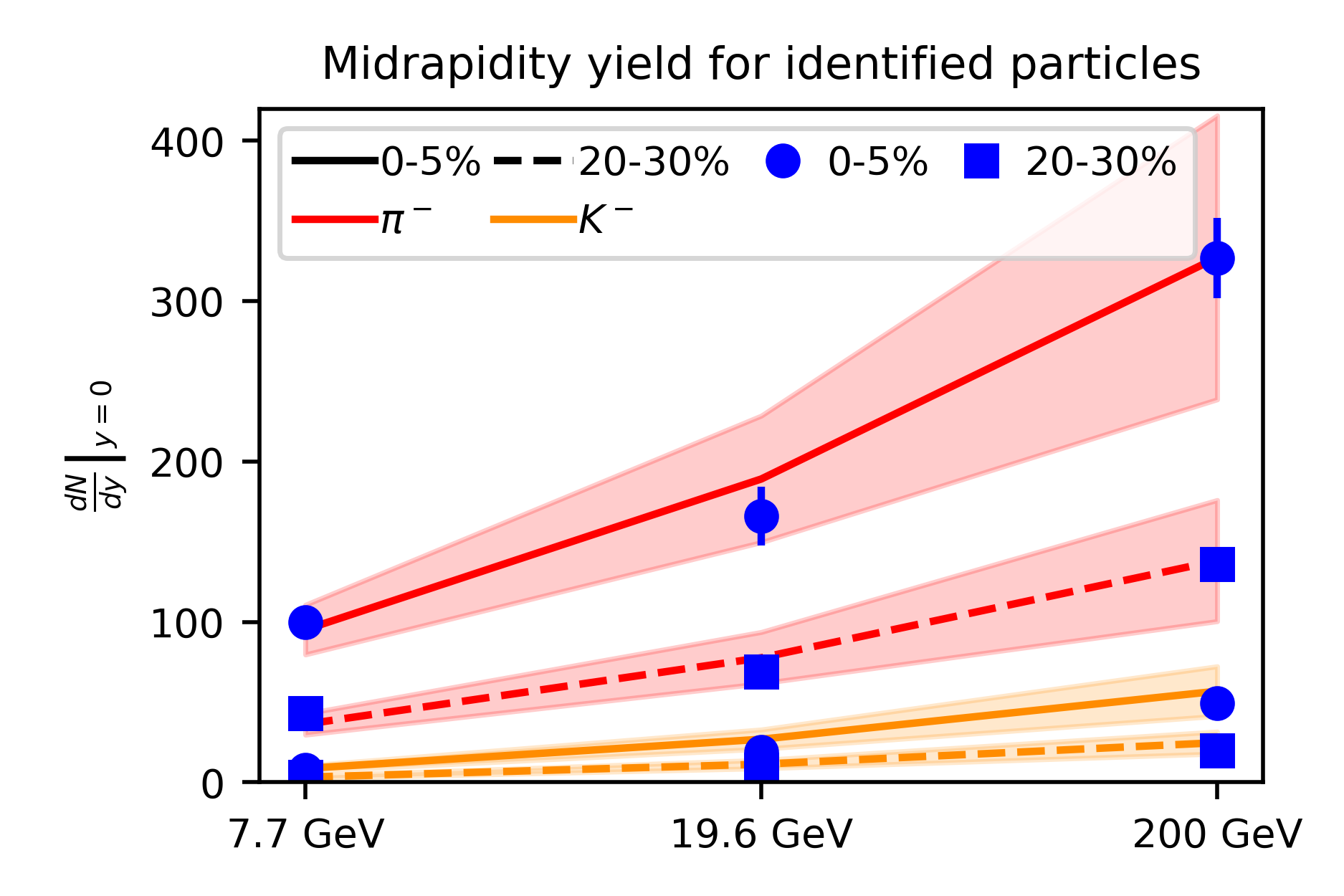}
    \includegraphics[width=\linewidth]{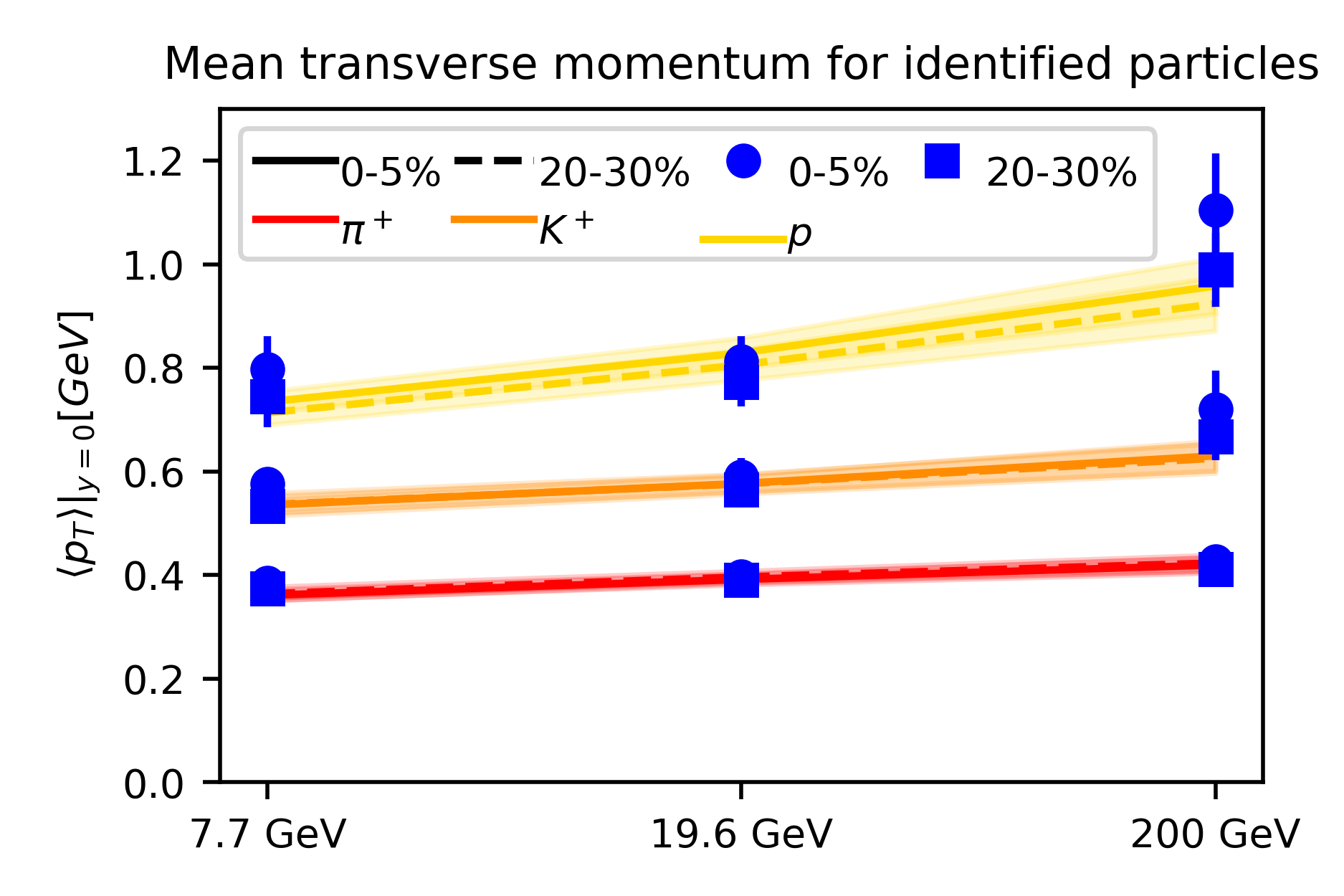}
    \includegraphics[width=\linewidth]{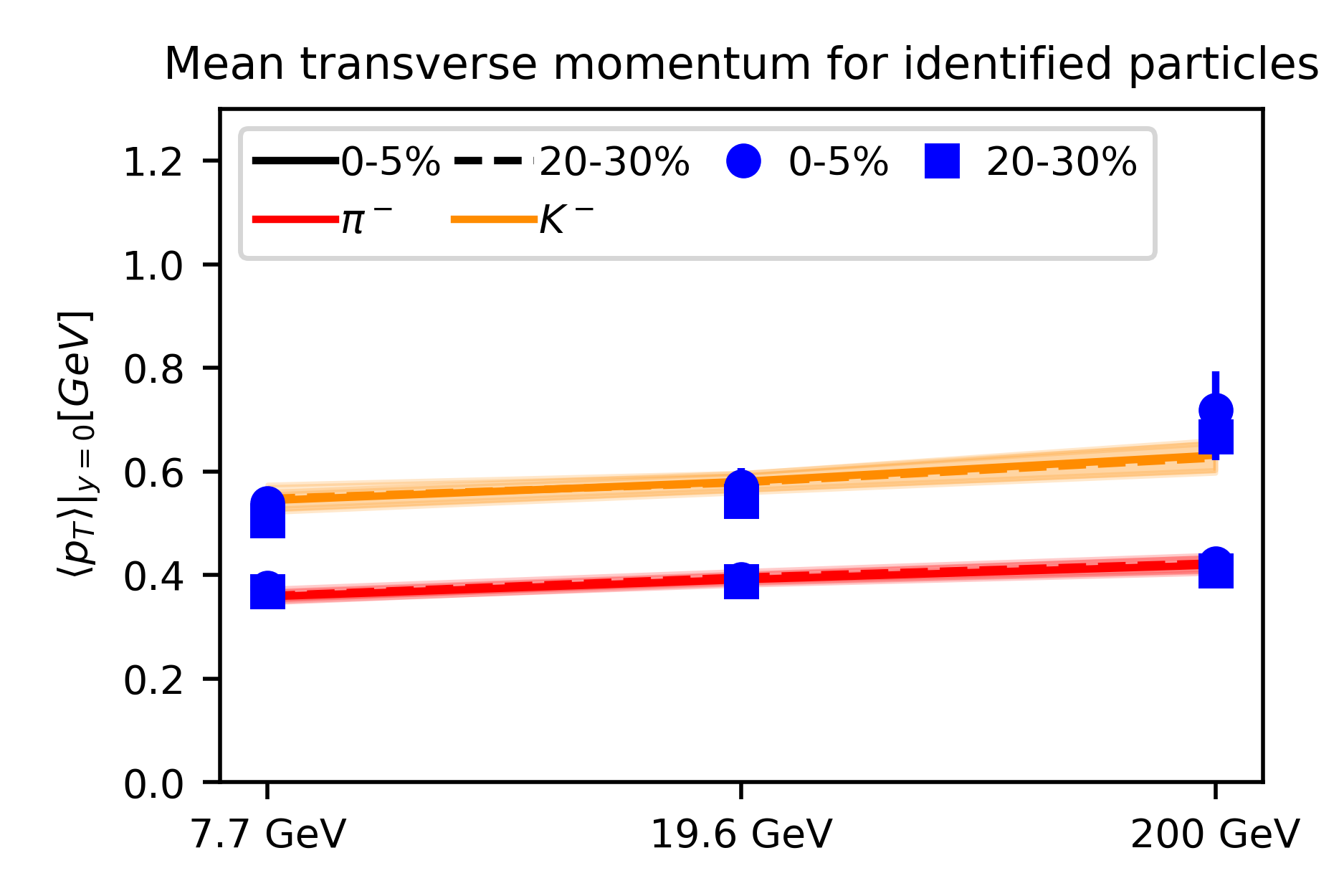}
     \caption{Comparison of predictions of observables from the posterior with experimental values for bulk observables.}
     \label{fig:datacomp_linear_bulk}
\end{figure}
\begin{figure*}
    \centering
    \includegraphics[width=0.45\linewidth]{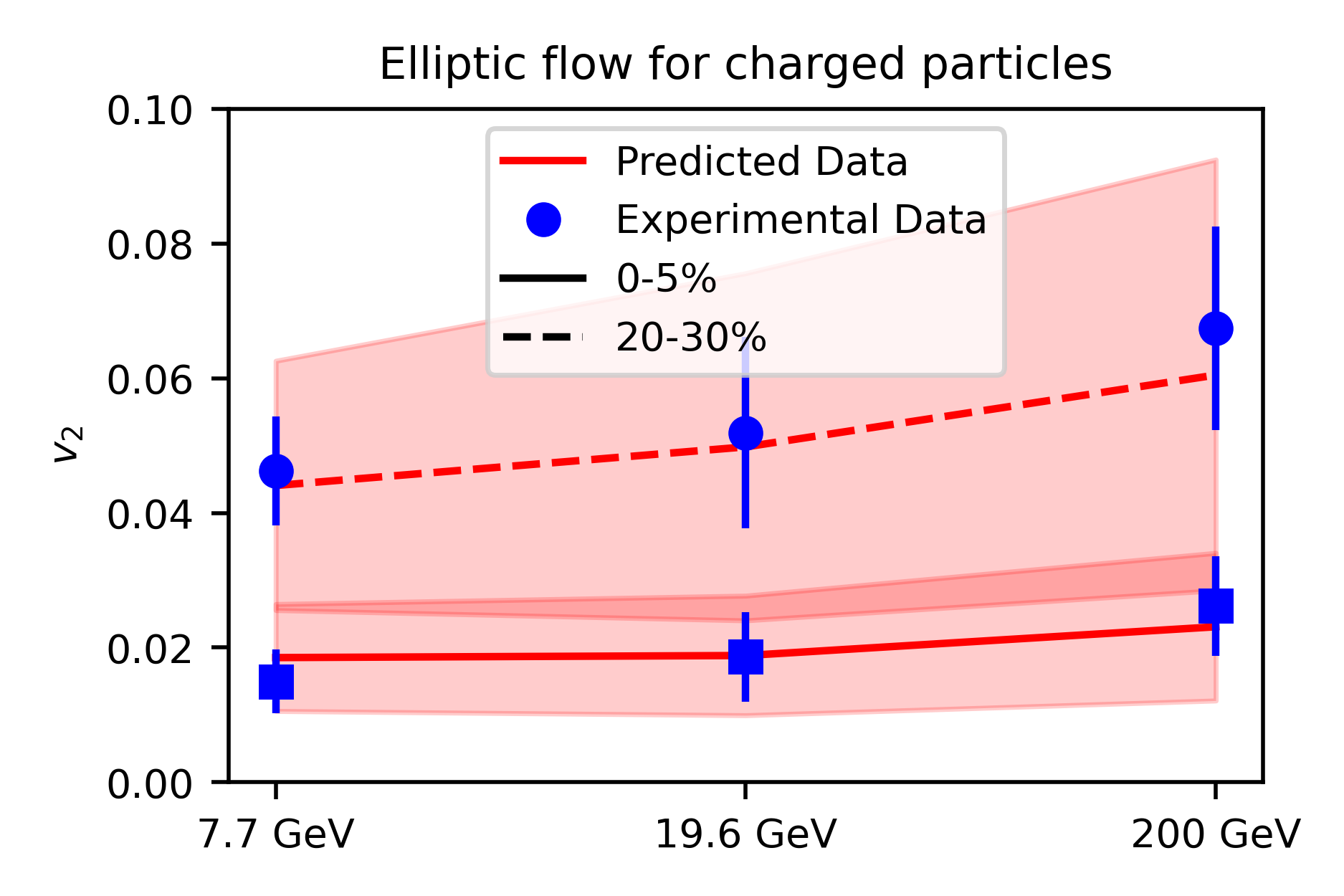}
    \includegraphics[width=0.45\linewidth]{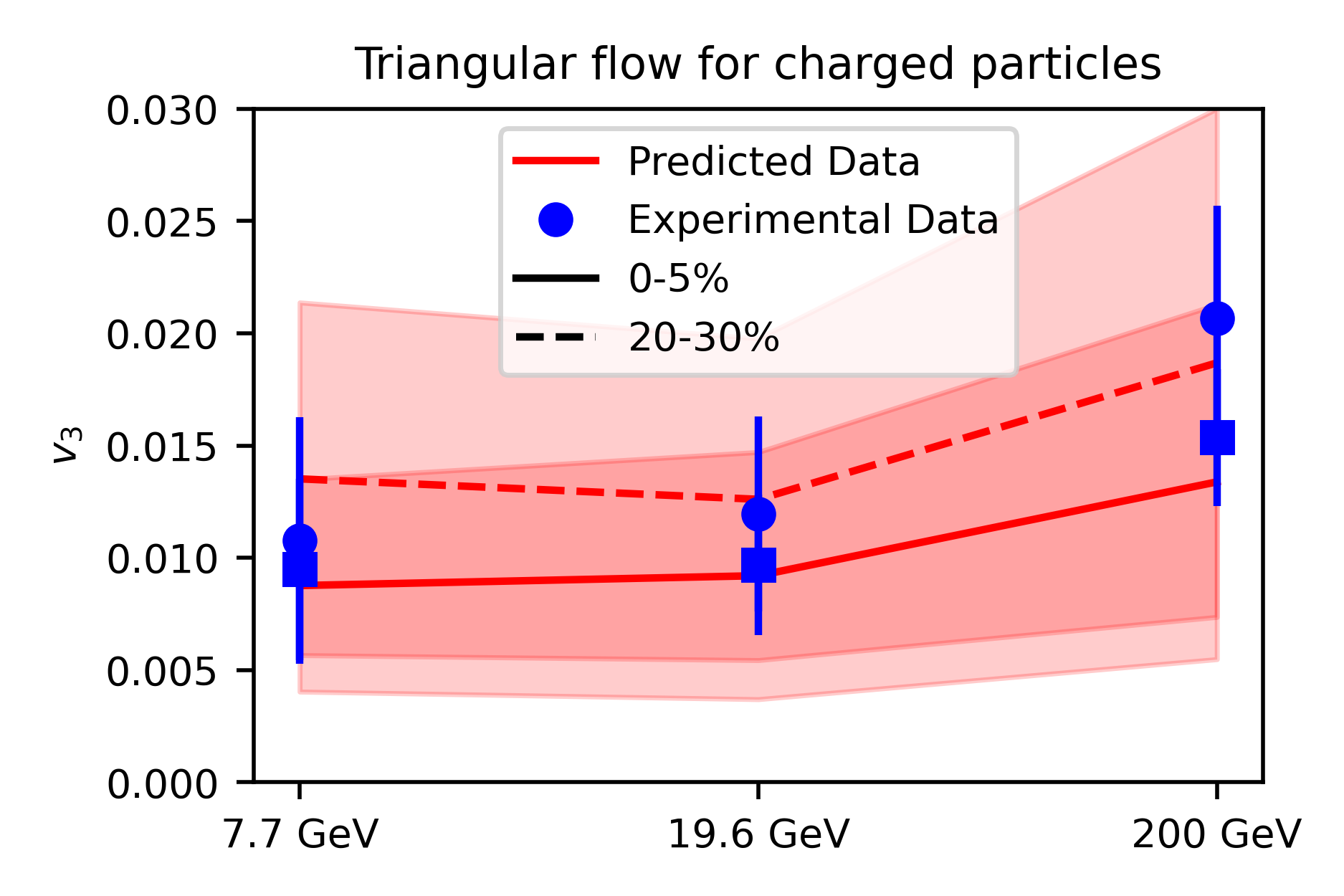}
    \includegraphics[width=0.45\linewidth]{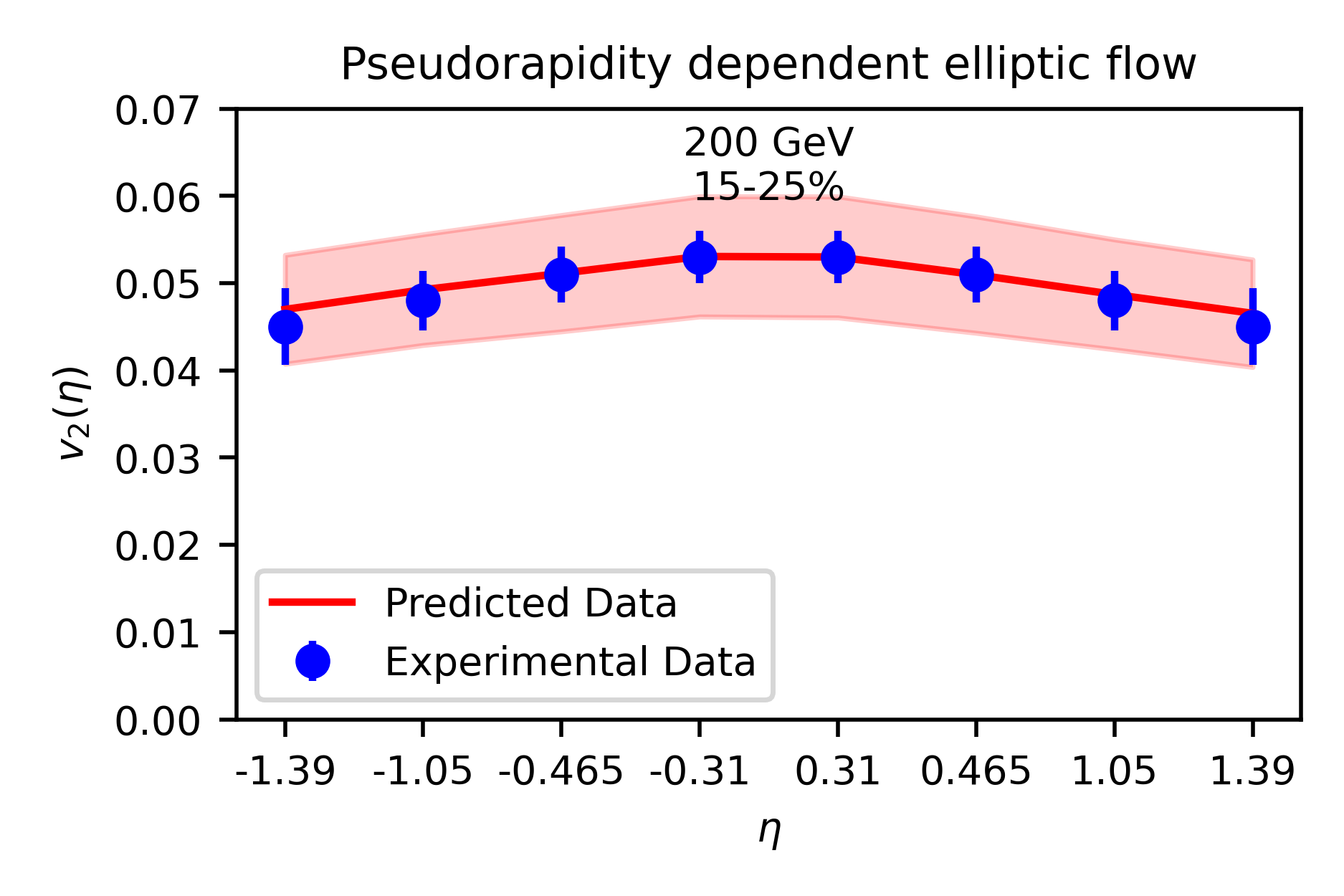}
    \caption{Comparison of predictions of observables from the posterior with experimental values for flow data.}
    \label{fig:datacomp_linear_flow}
\end{figure*}
\begin{figure*}
    \centering
     
    \includegraphics[width=\linewidth]{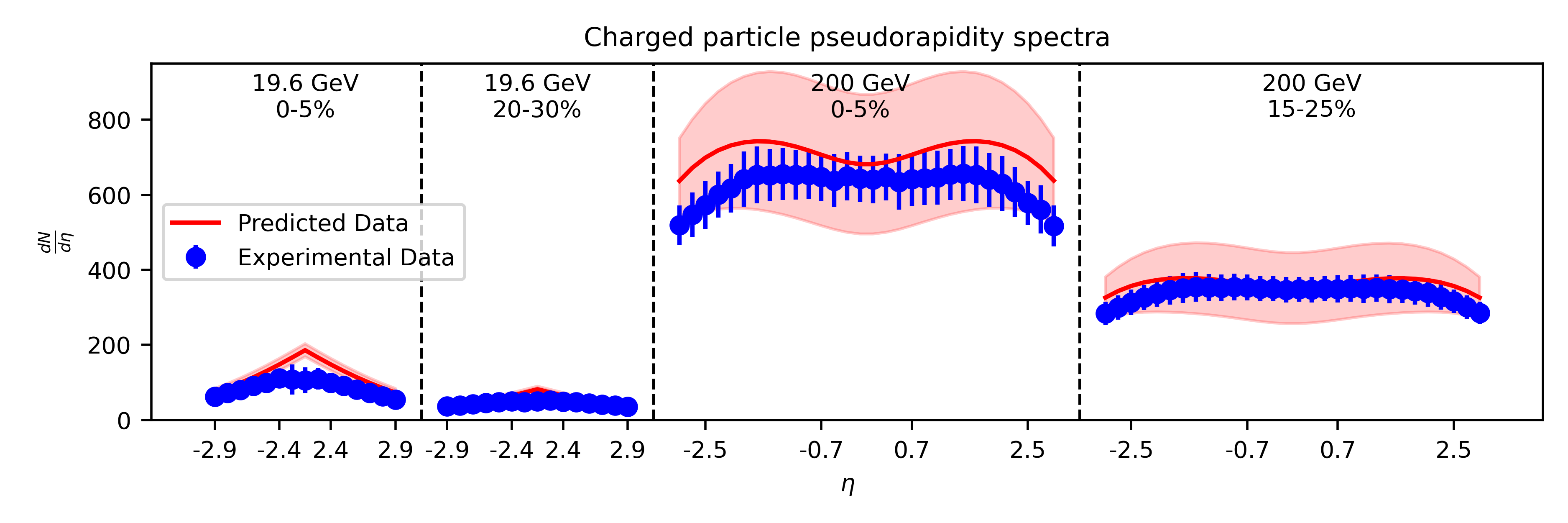}
     \caption{Comparison of predictions of observables from the posterior with experimental values for pseudorapidity yield.}
     \label{fig:datacomp_linear_eta}
\end{figure*}

\FloatBarrier
\bibliography{2025_bayes}
\end{document}